\documentclass[12pt]{iopart}
\usepackage{bm}
\usepackage{amssymb}
\usepackage{siunitx}
\usepackage{graphicx}
\usepackage{mathrsfs}  
\makeatletter
\DeclareRobustCommand{\text}{%
  \ifmmode\expandafter\text@\else\expandafter\mbox\fi}
\let\nfss@text\text
\def\text@#1{{\mathchoice
  {\textdef@\displaystyle\f@size{#1}}%
  {\textdef@\textstyle\f@size{#1}}%
  {\textdef@\textstyle\sf@size{#1}}%
  {\textdef@\textstyle \ssf@size{#1}}%
  \check@mathfonts
  }%
}
\def\textdef@#1#2#3{\hbox{{%
                    \everymath{#1}%
                    \let\f@size#2\selectfont
                    #3}}}
\makeatother

\begin{document}
\title[Theory of the Hall effect in three-dimensional metamaterials]{Theory of the Hall effect in three-dimensional metamaterials}

\author{Christian Kern$^{1, 2}$, Graeme W Milton$^3$, 
Muamer Kadic$^{1, 2, 4}$, and Martin Wegener$^{1, 2}$}

\address{$^1$ Institute of Applied Physics, Karlsruhe Institute of Technology (KIT),\\ 76128 Karlsruhe, Germany}
\address{$^2$ Institute of Nanotechnology, Karlsruhe Institute of Technology (KIT),\\ 76021 Karlsruhe, Germany}
\address{$^3$ Department of Mathematics, University of Utah, Salt Lake City, Utah 84112, USA}
\address{$^4$ Institut FEMTO-ST, UMR 6174, CNRS, Universit\'{e} de Bourgogne Franche-Comt\'{e},\\ 25000 Besan\c{c}on, France}

\ead{christian.kern@kit.edu}

\begin{abstract}
We apply homogenization theory to calculate the effective electric conductivity and Hall coefficient tensor of passive three-dimensionally periodic metamaterials subject to a weak external static homogeneous magnetic field. We not only allow for variations of the conductivity and the Hall coefficient of the constituent material(s) within the metamaterial unit cells, but also for spatial variations of the magnetic permeability. We present four results. First, our findings are consistent with previous numerical calculations for finite-size structures as well as with recent experiments. This provides a sound theoretical justification for describing such metamaterials in terms of effective material parameters. Second, we visualize the cofactor fields appearing in the homogenization integrals. Thereby, we identify those parts of the metamaterial structures which are critical for the observed effective metamaterial parameters, providing a unified view onto various previously introduced single-constituent/multiple-constituent and isotropic/anisotropic architectures, respectively. Third, we suggest a novel three-dimensional non-magnetic metamaterial architecture exhibiting a sign reversal of the effective isotropic Hall coefficient. It is conceptually distinct from the original chainmail-like geometry, for which the sign reversal is based on interlinked rings. Fourth, we discuss two examples for metamaterial architectures comprising magnetic materials: Yet another possibility to reverse the sign of the isotropic Hall coefficient and an approach to conceptually break previous bounds for the effective mobility.
\end{abstract}

\maketitle

%\tableofcontents

\section{Introduction}
Ordinary semiconductors such as silicon derive their importance in technology and every-day life from the possibility to adjust and tune their electric conductivity \cite{Yu05}. The large Hall coefficient of semiconductors also makes them ideal for Hall-effect based magnetic-field sensors \cite{Popovic04}. For example, such sensors underlie the compass apps in many modern mobile phones. High-mobility semiconductors are crucial for metrology standards \cite{QHE} based on the quantum Hall effect.

Metamaterials are rationally designed artificial composite materials that obtain their properties from structure rather than from chemistry alone. For optical \cite{Soukoulis11}, mechanical \cite{Bauer17}, acoustical \cite{Cummer16}, and thermal \cite{Alu14} properties, the metamaterial concept has led to effective material parameters going qualitatively and quantitatively beyond those of their constituent material(s). Particular emphasis has been laid on reversing the sign of effective material parameters with respect to its constituent(s), \textit{e.g.}, for the magnetic permeability \cite{Pendry99} and the refractive index \cite{Soukoulis07} at optical frequencies, the dynamic mass density \cite{Mei06}, the dynamic bulk modulus \cite{Fang06}, and the static poroelastic compressibility \cite{Gatt08,Qu17}. 

Much less work has been performed in regard to passive metamaterials with rationally designed electrical properties. At first sight, it appears as if not that much is possible: A negative effective DC electric conductivity would be in conflict with energy conservation and the second law of thermodynamics. In contrast, active structures comprising some sort of energy source can exhibit negative effective absolute mobility \cite{Ros05} and negative effective absolute resistance \cite{Yang13}. Moreover, the effective DC conductivity of a composite, \textit{e.g.}, made of two passive constituent materials A and B, is bounded by the non-negative DC conductivities of A and B, respectively \cite{Milton02}. 

The situation is distinct for the effective electric conductivity tensor and its elements in the presence of a static homogeneous magnetic field. Early mathematical work based on homogenization theory has shown that the effective Hall coefficient, which is directly linked to the off-diagonal elements of the effective conductivity tensor, can exhibit a sign reversal with respect to the constituent materials in three-dimensional (3D) structures \cite{Briane09}, but not in 2D structures for perpendicular magnetic fields \cite{Briane08}. Recent experiments based on 3D single-semiconductor-constituent structures \cite{Kern17} have brought these predictions to life. The parallel Hall effect \cite{Briane10, Kern15}, for which the Hall voltage drop is parallel to the magnetic field axis, has been observed in anisotropic 3D structures \cite{Kern17-2}. 

Such semiconductor metamaterials can be looked at from different angles. The mentioned original mathematical work \cite{Briane09} was based on homogenization theory \cite{Bensoussan78, Bergman83}. Additional idealizations allowed for an analytical treatment. Intuitively, one can see these metamaterials as networks of Hall-voltage sources, wired up in three dimensions by electrical connections \cite{Kern17-2}. Yet another approach is to calculate the behavior of a finite-size microstructure numerically, then consider it a black box, and map the result onto effective material parameters. While all of these viewpoints and approaches are valid in their own right, it is unsatisfactory that the early mathematical work \cite{Briane09} considered idealized microstructures that were actually quite different from the ones realized in experiments more recently \cite{Kern17}. This left it unclear, whether the physical mechanisms at work are the same.

In this paper, we therefore apply homogenization theory to the single-constituent/multiple-constituent and to the isotropic/anisotropic semiconductor architectures discussed so far and to new ones, thereby clarifying the relation between them. Our work also provides a sound a posteriori theoretical basis for the experimentally realized structures, by showing that their behavior can indeed be mapped onto effective conductivity tensors (or Hall coefficient tensors) in a well-defined manner. This aspect was recently questioned in comments by Mani \cite{Mani17} and Oswald \cite{Oswald18} (also see our responses \cite{Wegener17} and \cite{Kern18}, respectively). Furthermore, we expand theory to not only allow for spatial distributions of the conductivity and the Hall coefficient, but also of the magnetic permeability of the constituent material(s).

\section{Preliminaries}
In this paper, we study the static electric conductivity problem, 
\begin{equation}
\nabla \cdot \bm{j}=0,~~\bm{j}=\bm{\sigma}\left(\bm{b}\right)\bm{e},~~\nabla\times\bm{e}=0,
\label{elcond}
\end{equation}
where $\bm{j}$ is the electric current, $\bm{\sigma}\left(\bm{b}\right)$ is the magnetic field dependent electric conductivity tensor, and $\bm{e}$ is the electric-field vector. In terms of the electric potential $\phi$, with $-\nabla \phi=\bm{e}$, we obtain
\begin{equation}
\nabla \cdot \left(\bm{\sigma}\left(\bm{b}\right)\left(\nabla \phi\right) \right)=0.
\label{eqnpot}
\end{equation}
The electric conductivity tensor is constrained by fundamental considerations. Onsager's principle \cite{Onsager31, Onsager31_2, LandauStat, LandauCont} implies that
\begin{equation}
\bm{\sigma}\left(\bm{b}\right) = \bm{\sigma}\left(-\bm{b}\right)^{\intercal}.
\label{Onsager}
\end{equation}
We are interested in the regime of small magnetic fields. Hence, we expand the conductivity tensor up to the first order in the magnetic field, $\bm{\sigma}\left(\bm{b}\right)=\bm{\sigma}_0+\bm{\sigma}_1\left(\bm{b}\right)$, where $\bm{\sigma}_0$ is the zero magnetic field conductivity tensor and $\bm{\sigma}_1\left(\bm{b}\right)$ is linear in $\bm{b}$.
Equation\,(\ref{Onsager}) implies that $\bm{\sigma}_0^{\intercal}=\bm{\sigma}_0$ and $\bm{\sigma}_1\left(\bm{b}\right)^{\intercal}=-\bm{\sigma}_1\left(\bm{b}\right)$. As $\bm{\sigma}_1\left(\bm{b}\right)$ is linear in the magnetic field and antisymmetric, we have that \cite{LandauCont, Briane09}
\begin{equation*}
\bm{\sigma}_1\left(\bm{b}\right) = \mathscr{E}\left(\bm{S}\bm{b}\right),
\end{equation*}
where $\bm{S}\in\mathbb{R}^{3\times 3}$ is some matrix, which we refer to as the \textit{S}-matrix, and $\mathscr{E}$ is the Levi-Civita tensor
\begin{equation*}
\mathscr{E}: \mathbb{R}^3 \rightarrow \mathbb{R}^{3\times3},~\mathscr{E}: \bm{x} \mapsto \mathscr{E} \left(\bm{x} \right),
\end{equation*}
\begin{equation*}
\left(\mathscr{E}\left(\bm{x} \right)\right)_{ij}=\sum_{k=0}^{3}\epsilon_{ijk}x_k,
\end{equation*}
\begin{equation*}
\epsilon_{ijk} = \left\{
\begin{tabular}{ll} 
$1$ & if $(i,j,k)$ is an even permutation of $(1,2,3)$  \\
$-1$ & if $(i,j,k)$ is an odd permutation of $(1,2,3)$
\end{tabular}\right.
\end{equation*}
The same considerations apply to the resistivity tensor, and hence, we obtain
\begin{equation*}
\bm{\rho} = \bm{\rho}_0+\mathscr{E} \left(\bm{A}_{\text{H}}\bm{b}\right),
\end{equation*}
\begin{equation*}
\text{and }\bm{\sigma} = \bm{\sigma}_0+\mathscr{E} \left(\bm{S}\bm{b}\right),
\end{equation*}
where we have introduced the Hall matrix, $\bm{A}_{\text{H}}$.
Using $\bm{\sigma}=\bm{\rho}^{-1}$, it follows that, up to the first order in the magnetic field, the Hall matrix and the corresponding \textit{S}-matrix are connected \cite{Briane09} \textit{via} 
\begin{equation}
\bm{S}=-\text{Cof}\left(\bm{\sigma}_0\right)\bm{A}_{\text{H}}.
\label{HallmatSmat}
\end{equation}
Here, for any matrix $\bm{A}\in\mathbb{R}^{3\times 3}$, $\text{Cof}\left(\bm{A}\right)$ denotes the corresponding cofactor matrix, which is given by
\begin{equation*}
\left(\text{Cof}\left(\bm{A}\right)\right)_{ij}=(-1)^{(i+j)}\bm{M}(i,j),
\end{equation*}
where $\bm{M}(i,j)$ is the $\left(i,j\right)$-minor of $\bm{A}$, \textit{i.e.}, the determinant of the submatrix formed by deleting the $i$-th row and $j$-th column.\\

In general, any antisymmetric three-dimensional rank-2 tensor $\bm{M},~\bm{M}^{\intercal}=-\bm{M}$ is dual to a pseudovector $\bm{m}\in\mathbb{R}^3$ with $\bm{M}=\mathscr{E}\left(\bm{m}\right)$.
Then, the matrix product $\bm{M}\bm{x}$, for any three-dimensional vector $\bm{x}\in\mathbb{R}^3$ is simply the cross product $\bm{m}\times\bm{x}$. Here, $\mathscr{E}\left(\bm{A}_{\text{H}}\bm{b}\right)$ is dual to $\bm{A}_{\text{H}}\bm{b}$.
Thus, we can write the constitutive equation in the following form
\begin{equation}
\bm{e}=\bm{\rho}_0\bm{j}+\left(\bm{A}_{\text{H}}\bm{b}\right)\times\bm{j}
\label{constcr}
\end{equation}
\begin{equation*}
\text {or }\bm{j}=\bm{\sigma}_0\bm{e}+\left(\bm{S}\bm{b}\right)\times\bm{e}.
\end{equation*}
Note that, as $\bm{A}_{\text{H}}\bm{b}$ is a pseudovector (because $\bm{\rho}$ is a tensor) and $\bm{b}$ is a pseudovector as well,  $\bm{A}_{\text{H}}$ is a tensor. The same holds true for $\bm{S}$ \cite{LandauCont}.

\noindent
In the simple case of an isotropic conductor, we have $\bm{\sigma}_{0}=\sigma_{0}\bm{I}$, $\bm{A}_{\text{H}}=A_{\text{H}}\bm{I}$, and $\bm{S}=-\sigma_0^2A_{\text{H}}\bm{I}$,
where $A_{\text{H}}$ is the Hall coefficient and $\bm{I}$ is the identity matrix.\\

In the following, we consider the Hall effect in composites. As we will see, one has a lot of freedom in tailoring the individual elements of the effective Hall matrix by structure, even in simple single-constituent porous structures. 

The theory of composites describes materials that are made from one or more constituents and which are structured on a very small length scale. On this length scale, the physics is described by one or several partial differential equations (in our case equation\,(\ref{eqnpot})). The very fine structuring is reflected in the rapid oscillation of the coefficients of the equation(s) (in our case the electric conductivity $\bm{\sigma}$). On the macroscopic scale, the physics can very often be described by the same differential equation(s), however with smoothly varying or, depending on the problem, even constant coefficients. Roughly speaking, if we structure a material finer and finer, it will, on a large length scale, behave like a homogeneous material with different properties (see, \textit{e.g.}, \cite{Bensoussan78}). These properties are termed effective properties and the finely-structured material is said to be an effective material. Here, we restrict ourselves to the case of periodic composites. We briefly mention that this problem can be treated in a mathematical rigorous fashion. In the framework of H-convergence introduced by Murat and Tartar it is described by the convergence of sequences of not-necessarily periodic non-symmetric tensor fields \cite{Murat97}. The initial theoretical publication giving the first example of a material with a sign-inverted effective Hall coefficient was given in the language of H-convergence \cite{Briane09}.

Naively, one would think that the effective properties of a composite lie somewhere in between the properties of the constituents of the material. An astonishing and fascinating result of the theory of composites is however, that the properties of the effective material can be very much different from the properties of each of the constituents. In the case of the Hall effect, the effective Hall coefficient can be sign-inverted. Furthermore, it is possible to tailor the different elements of the effective Hall matrix.
In the following, we derive an expression for the effective Hall matrix from perturbation theory to the first order in the magnetic field.

In a composite, we have to distinguish between the microscopic fields, here the microscopic electric field $\bm{e}$ and the microscopic electric current density $\bm{j}$, solving the partial differential equation on the small length scale
and the macroscopic fields solving the homogenized equation with the effective coefficients. The macroscopic fields vary on a length scale that is large compared to the unit cell of periodicity. They can be obtained by averaging the microscopic fields over a region in space intermediate between the lattice constant and macroscopic characteristic lengths such as the size of the macroscopic body or the length scale of variations of the electrostatic potential applied at the boundary of this body. Due to the separation of length scales, we can assume that the macroscopic fields are constant over a single unit cell. Then, any microscopic field $\bm{e}$ with an average value $\langle \bm{e}\rangle$, which is the corresponding macroscopic field, is given by 
\begin{equation}
\bm{e}= -\left(\nabla \bm{\Phi}\right)\langle \bm{e}\rangle,~\left(\nabla \bm{\Phi}\right)_{ij} = \frac{\partial \phi_j}{\partial x_i}.
\label{fiemicr}
\end{equation}
Here, $\bm{\phi}$ is the vector-valued electric potential $\bm{\Phi}=\left(\phi_1,\phi_2,\phi_3\right)^{\intercal}$ which solves
\begin{equation}
\nabla \cdot \left(\bm{\sigma}\nabla \bm{\Phi}\right)=0
\label{vecvalelpot}
\end{equation}
and which is subject to the boundary condition that $\bm{\Phi}\left(\bm{y}\right)+\bm{y}$ is invariant with respect to translations by integer multiples of one unit cell. The three fields $\phi_1$, $\phi_2$, and $\phi_3$ are the three microscopic electric potentials corresponding to the average electric field pointing along each of the three axes. For an arbitrary direction, the microscopic electric potential is given by a linear combination of these fields (see also equation\,(\ref{fiemicr})). The electric field corresponding to the vector-valued electric potential $\bm{E}=-\nabla\bm{\Phi}$ is matrix-valued and normalized, $\left\langle \bm{E} \right\rangle=\bm{I}$. We note that the sign is often chosen differently. For example, in \cite{Briane09}, it is chosen such that $\left\langle \nabla\bm{\Phi} \right\rangle=\bm{I}$.

\noindent
On the macroscopic length scale, the constitutive equation reads as 
\begin{equation*}
\left\langle \bm{j}\right\rangle = \bm{\sigma}^* \left\langle \bm{e}\right\rangle.
\end{equation*}

\noindent
Hence, the effective conductivity tensor can be obtained from
\begin{equation}
\bm{\sigma}^* = \bm{\sigma}^* \left\langle \bm{E}\right\rangle = -\left\langle \bm{\sigma}\nabla\bm{\Phi}\right\rangle.
\label{constitmacr}
\end{equation}

In the following, we treat the problem for finite magnetic fields as a perturbation to the zero magnetic-field problem. In this limit, the influence of the magnetic field on the conductivity tensor is small. Then, it is sufficient to solve equation\,(\ref{vecvalelpot}) for zero magnetic field $\left(\bm{\sigma} = \bm{\sigma}_0\right)$. The zero magnetic field effective conductivity $\bm{\sigma}_0^*$ can be obtained from equation\,(\ref{constitmacr}). To obtain the magnetic field dependent effective conductivity---and the effective Hall matrix---we use the following perturbative expression (see, \textit{e.g.} Chapter 16 in \cite{Milton02}),
\begin{equation}
\langle \bm{e}'\rangle \cdot \delta\bm{\sigma}^*\langle \bm{e}\rangle = \langle \bm{e}'\cdot\left(\delta\bm{\sigma}\right)\bm{e}\rangle.
\label{eqge}
\end{equation}
with $\delta\bm{\sigma} = \mathscr{E} \left(\bm{S}\bm{b} \right)$ and $\delta\bm{\sigma}^* = \mathscr{E} \left(\bm{S}^*\bm{b} \right)$. Here, $\bm{e}$ and $\bm{e}'$ are solutions to equation\,(\ref{elcond}). In general, $\bm{e}'$ is a solution to the adjoint problem. Here however, we make use of the fact that $\bm{\sigma}_0$ is symmetric. As this equation has to hold for all solutions, we can substitute the matrix-valued electric field solution into equation\,(\ref{eqge})  
\begin{equation}
\langle \bm{E}\rangle \cdot \delta\bm{\mathcal{\sigma}}^*\langle \bm{E}\rangle = \langle \bm{E}\cdot\left(\delta\bm{\sigma}\right)\bm{E}\rangle.%\label{eqge}
\end{equation}
As the average matrix-valued electric field is normalized, $\left\langle\bm{E}\right\rangle=\bm{I}$, the left-hand side of equation\,(3) can be evaluated easily. 
Then
\begin{equation*}
\delta\bm{\sigma}^*=\left\langle \left(\nabla\bm{\Phi}\right)^{\intercal}\delta\bm{\sigma}\left(\nabla\bm{\Phi}\right)\right\rangle\\
\Rightarrow\mathscr{E}\left(\bm{S}\bm{b}\right)^*=\left\langle\mathscr{E}\left(\text{Cof}\left(\nabla\bm{\Phi}\right)^{\intercal}\bm{S}\bm{b}\right)\right\rangle,
\end{equation*}
and because this has to hold for all values of the magnetic field one gets
\begin{equation*}
\bm{S}^*=\left\langle\text{Cof}\left(\nabla\bm{\Phi}\right)^{\intercal}\bm{S}\right\rangle.
\end{equation*}
Using equation\,(\ref{HallmatSmat}), one arrives at the following expression for the effective Hall matrix \cite{Briane09, Briane10}
\begin{equation}
\left\langle\text{Cof}\left(\bm{\sigma}_0\nabla\bm{\Phi}\right)^{\intercal} \bm{A}_{\text{H}}\right\rangle=\text{Cof}\left(\bm{\sigma}_0^*\right)\bm{A}_{\text{H}}^*.
\label{effHallmat}
\end{equation}
Throughout this paper, we use this expression to determine the effective Hall matrix of various microstructures. Depending on the symmetry of the structure, simplifications may apply. We emphasize that (\ref{effHallmat}) holds true in the limit of small magnetic fields. Few studies have gone beyond this limit \cite{Bergman94, Tornow96, Bergman17}.

In the numerical calculations, we solve equation\,(\ref{vecvalelpot}) with $\bm{\sigma}=\bm{\sigma}_0$ using COMSOL Multiphysics, more precisely using its Electric Currents module. The equation is solved separately for all three directions, \textit{i.e.}, for $\phi_1$, $\phi_2$ and $\phi_3$. We consider cubic unit cells with lattice constant $a$. The periodic boundary conditions for the electric potential are implemented using the corresponding built-in function. The periodicity with a potential drop is manually implemented using a weak contribution. As the electric potential is defined only up to a constant, we fix the potential at a single point in the structure.

The boundary conditions can be simplified significantly, depending on the symmetry of the structure. In the case of mirror symmetry with respect to the $xy$-, $yz$-, and $xz$-plane (passing through the center of the unit cell), we obtain the following boundary conditions for $\phi_1$, $\phi_1(-a/2,x,y) = c,~\phi_1(a/2,x,y)=c+a$ and all other boundaries are insulating, where $c$ is a constant that can be chosen arbitrarily. The boundary conditions for $\phi_2$ and $\phi_3$ are analogous. These boundary conditions facilitate the calculations for certain microstructures, including the chainmail-inspired geometry.

Once we know the vector-valued electric potential, we can determine the cofactor matrix of the corresponding current field easily and determine the effective Hall coefficient from a volume average. First, we use equation\,(\ref{constitmacr}) to determine the zero magnetic-field effective conductivity $\bm{\sigma}_0^*$. Second, we use equation\,(\ref{effHallmat}) to determine the effective Hall matrix.

In principle, one could also solve equation\,(\ref{vecvalelpot}) with the full magnetic field dependent conductivity $\bm{\sigma}\left(\bm{b} \right)$ instead of $\bm{\sigma}_0$. Such a calculation would be somewhat more complicated, but more importantly, it does not offer any deeper understanding. In contrast, equation\,(\ref{effHallmat}) can serve as a tool for the design of microstructures with desired properties and can be an intuitive access to the problem.

We note that many results of homogenization theory require that $\bm{\sigma}_0$ is bounded and coercive in the sense that there exists some constant $\alpha>0$ such that $\bm{\sigma}_0(x) \geq \alpha \bm{I}$ for all $x$ (where the matrix inequality holds in the sense of quadratic forms). Hence, in order to treat porous structures, as considered below, mathematically rigorously, one has to add a very weakly conducting surrounding medium. This does not affect the results presented here. In the more general case, treated by Camar-Eddine and Seppecher \cite{camareddine01, camareddine02}, one can, \textit{e.g.}, obtain non-local behavior. 

\section{Isotropic structures and sign-inversion of the Hall coefficient}
In the following we consider an isotropic conductor. In this case, equation (\ref{effHallmat}) reduces to the following formula for the effective Hall coefficient, 
\begin{equation*}
\left\langle\left(J_{11} J_{22}-J_{21} J_{12}\right) A_{\text{H}}\right\rangle=\left(\sigma_0^*\right)^2 A_{\text{H}}^*.
\end{equation*}
Here, $\bm{J}$ is the current field associated with $\bm{\Phi}$ and $J_{11} J_{22}-J_{21} J_{12}=\left(\text{Cof}\left(\bm{J}\right)\right)_{33}=\left(\text{Cof}\left(\bm{\sigma}\nabla\bm{\Phi}\right)\right)_{33}$. Of course, one could equivalently consider any other of the three diagonal cofactors. The expression is somewhat simpler, if, instead of the electric field, the current is normalized, $\left\langle \tilde{\bm{J}} \right\rangle=\bm{I}$. Then,
\begin{equation*}
A_{\text{H}}^*=\left\langle \left(\tilde{J}_{11} \tilde{J}_{22}-\tilde{J}_{21} \tilde{J}_{12}\right) A_{\text{H}}(\bm{x})\right\rangle.
\end{equation*}
This result was first obtained by Bergman \cite{Bergman83}.

\begin{figure}[h]
	\includegraphics[width=16cm]{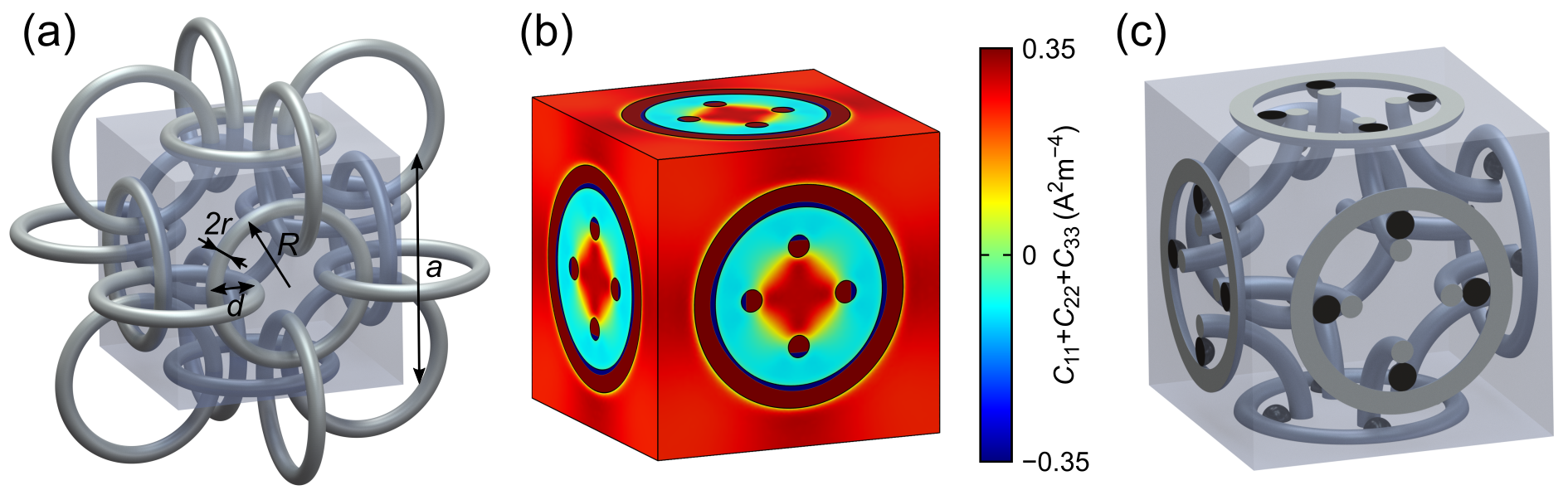}
	\caption{(a) A single extended unit cell of the cubic structure comprised of interlinked highly conductive tori embedded in a weakly conducting medium introduced in \cite{Briane09}. (b) Numerical calculation of the trace of the cofactor matrix, $C_{11}+C_{22}+C_{33}$, of the matrix-valued electric current for the structure shown in (a). Note that in between the tori, the trace turns negative. An effective material exhibiting a sign-inversion of the Hall coefficient can be obtained by placing a material with nonzero Hall coefficient there and choosing the Hall coefficient to be zero everywhere else \cite{Briane09, Kadic15}. Parameters are defined as in \cite{Kern17}, $R=36\,\si{\micro\meter}$, $d=-18\,\si{\micro\meter}$, $r=4\,\si{\micro\meter}$, $a=108\,\si{\micro\meter}$, $\sigma_0^{\text{Tori}}=200\,\si{\siemens\per\meter}$ and $\sigma_0^{\text{Surr.}}=0.2\,\si{\siemens\per\meter}$, where $\sigma_0^{\text{Tori}}$ is the conductivity of the tori and $\sigma_0^{\text{Surr.}}$ is the conductivity of the surrounding material. (c) A single unit cell of such an effective material \cite{Kadic15, Kern17}. The only parts with a nonzero Hall coefficient are the small black spheres placed in between the intertwined tori.}
\end{figure}
In isotropic structures, it is possible to invert the sign of the Hall coefficient. The first example of such a microstructure was based on a related microstructure exhibiting a local sign-inversion of the corrector's determinant, $\text{det}\left(\nabla \bm{\Phi}\right)$ \cite{Briane04}. One of the main questions of that paper was, whether any effective tensor of an arbitrary microstructure can be realized with a hierarchical laminate and a given set of constituents. The authors were able to show that in a three-dimensional hierarchical laminate the corrector's determinant is positive almost everywhere. Previously, Alessandrini and Nesi had shown that this holds for all two-dimensional microstructures \cite{Alessandrini01}, which implies that there is no effective material exhibiting a sign-inversion of the Hall coefficient in two dimensions with the magnetic field perpendicular to the plane of conduction \cite{Briane08}. The authors then presented a three-dimensional effective material made from linear chains of interlinked tori, in which the determinant turns negative locally, more precisely in between the interlinked tori. Therefore, following \cite{Briane08}, the range of realizable properties is larger than that of hierarchical laminates. The authors' linear-chain microstructure has been extended into the chainmail-inspired three-dimensional cubic effective material exhibiting a sign-inversion of the effective Hall coefficient.
The corresponding structure is shown in figure\,1(a). Highly conducting interlinked tori are embedded in weakly conducting surrounding medium. The constituents are electrically isotropic.
\begin{figure}[h]
	\includegraphics[width=6cm]{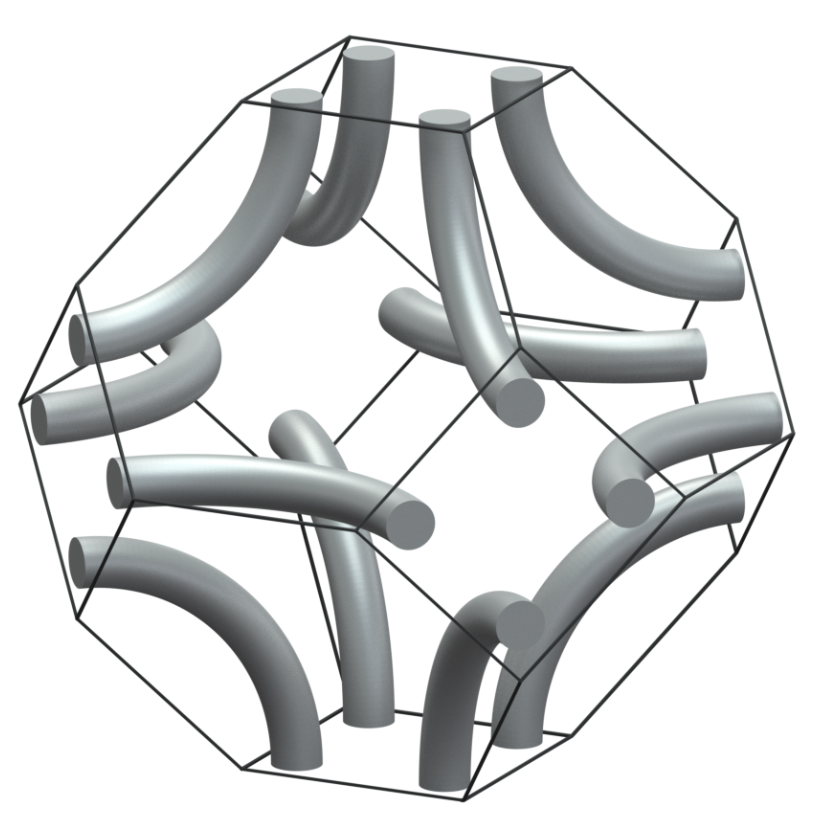}
	\caption{The Wigner-Seitz primitive unit cell of the body-centered cubic based arrangement of interlinked tori.}
\end{figure}
We note that this structure can be described by a body-centered cubic lattice with the three-atomic basis
\begin{equation*}
 \text{T}_x:\left(a/2,0,0\right)^{\intercal},~\text{T}_y:\left(0,a/2,0\right)^{\intercal},~\text{T}_z:\left(0,0,a/2\right)^{\intercal},
\end{equation*}
where each atom, $\text{T}_x$, $\text{T}_y$ and $\text{T}_z$, is a torus, the index corresponding to its axis. A corresponding Wigner-Seitz unit cell is shown in figure\,2.

\begin{figure}[h]
	\includegraphics[width=8cm]{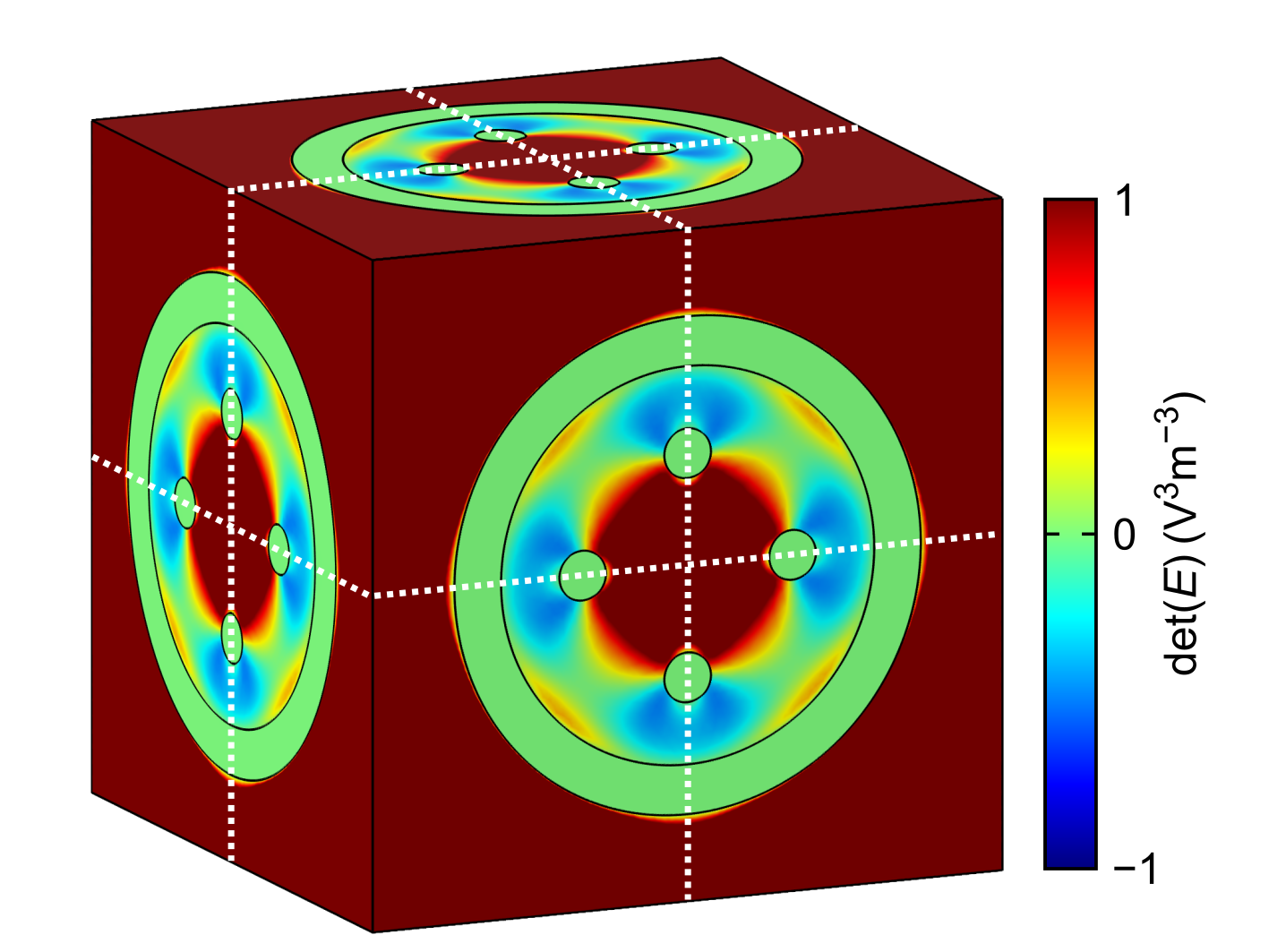}
	\caption{Numerically evaluated determinant of the matrix-valued electric field for the structure shown in figure\,1(a). Along the dashed white lines, along which the symmetry of the structure is high, the sign of the corresponding cofactor is given by the sign of the determinant. In between the interlinked tori, the determinant turns negative. Parameters are as in figure\,(1).}
\end{figure}
The results of a corresponding numerical calculation of the sum of the diagonal cofactors of the matrix-valued current density are shown in figure\,1(b). Notably, this sum turns negative in between the intertwined tori. There is a close connection between the diagonal cofactors of the matrix-valued current density and the determinant of the matrix-valued electric field (the corrector). First we note that, as the constituents are isotropic, the cofactor of the matrix-valued current density is, up to a constant factor, simply the cofactor of the matrix-valued electric field. Along certain lines of high symmetry, indicated by white dashed lines in figure\,3, the sign of determinant of the matrix-valued electric field gives the sign of the corresponding diagonal cofactor. Consider the line defined by $y=0$ and $z=a/2$. Here, the structure has mirror symmetry with respect to two planes, perpendicular to the $y$- and $z$-direction, respectively. This symmetry implies that $\bm{E}$ is diagonal,
\begin{equation*}
\bm{E}(x,0,a/2)=\left(\begin{array}{ccc} E_{11} &  0 & 0\\  0  & E_{22} & 0\\ 0 & 0 & E_{33}\\ \end{array}\right)(x,0,a/2)
\end{equation*}
and we obtain
\begin{equation*}
\left(\text{Cof}\left(\bm{E}\right)\right)_{33}=E_{11}E_{22} \text{ and } \det{\left(\bm{E}\right)}=E_{11}E_{22}E_{33}.
\end{equation*}
Along this high-symmetry line, $E_{22}$ and $E_{33}$ are positive while $E_{11}$ turns negative in between the intertwined tori (see \cite{Briane04} for a formal derivation). Therefore, the cofactor as well as the determinant turn negative there as well. By placing small spheres with a finite Hall coefficient there, one obtains an effective material, shown in figure\,1(c), with a sign-inverted Hall coefficient \cite{Briane09}.

We have shown previously that one can obtain a sign-inversion of the effective Hall coefficient in a similar single-constituent porous material \cite{Kadic15}. We omit the surrounding material and replace the spheres by cylinders made from the same material as the tori. The conductivity and the Hall coefficient are the same in all parts of the structure. The results of a numerical calculation of the cofactor $C_{33}=\text{Cof}\left(\bm{\sigma}_0\nabla\bm{\Phi}\right)_{33}$, corresponding to a magnetic field along the $z$-direction, are shown in figure\,4(a). The sign of the effective Hall coefficient is inverted as the overall volume average is negative. In most parts of the structure however, $C_{33}$ is close to zero, meaning that the local Hall voltages there do not enter into the global effect. These parts rather serve as interconnections, wiring up the regions in which $C_{33}$ is large, which can be seen as local Hall elements. These are mainly the regions where the cylinders and tori intersect. This finding is in good agreement with our previous intuitive explanation \cite{Kern17}. Consider a torus in the $xy$-plane, a magnetic field along $\hat{\bm{z}}$, and a current flowing in the $x$-direction. Local Hall voltages will appear in the torus, which are picked up by the cylinders connecting them to the tori in the $yz$-plane. Hence, we expect that $C_{33}$ takes large values in the torus close to the cylinders. In figure\,4(b), the sum of the diagonal cofactors, reflecting the symmetry of the structure, is shown. The sign-inversion is a result of the voltage being picked-up and the current being injected on the inner sides of the tori. This configuration of intertwined tori corresponds to a negative value of the distance parameter $d$. A positive value of $d$ corresponds to non-intertwined tori, see figure\,4(c). In this case, the overall average and hence, the effective Hall coefficient is positive.

\begin{figure}[h]
	\includegraphics[width=16cm]{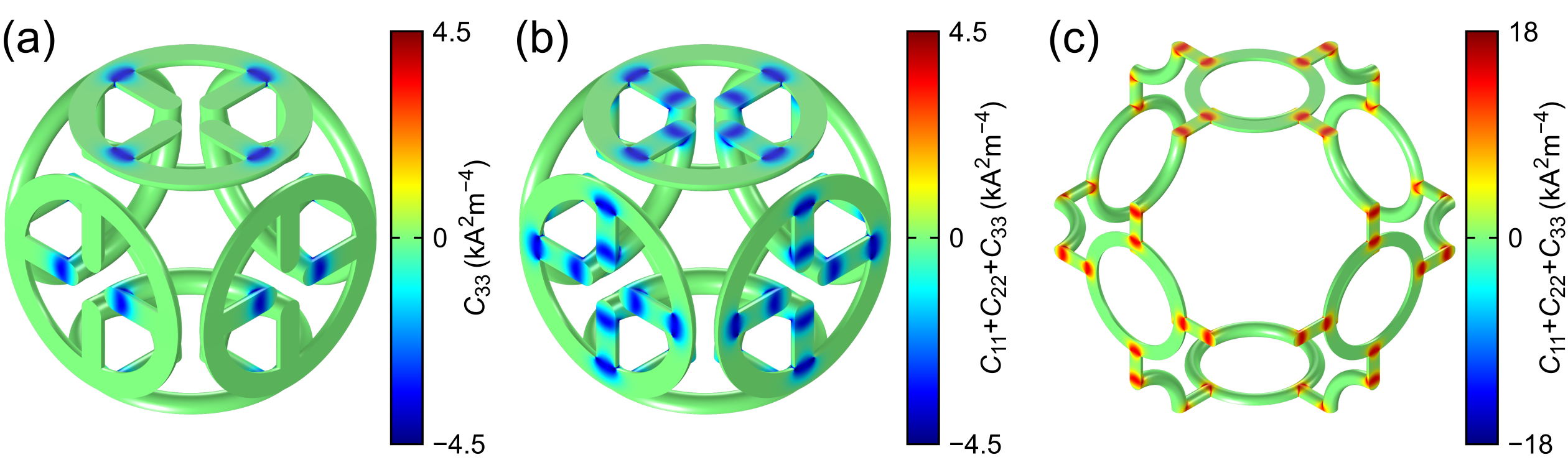}
	\caption{Numerical calculations of (a) one of the three diagonal cofactors, $C_{33}$ corresponding to a magnetic field along the $z$-direction, and (b) the trace of the cofactor matrix of the matrix-valued electric current for the chainmail-inspired sign-inversion structure \cite{Kadic15}, with a negative distance parameters, $d=-34\,\si{\micro\meter}$ corresponding to intertwined tori. In most parts of the structure, the moduli of the cofactors are small. The effective Hall coefficient is mainly determined by the regions where the tori and cylinders intersect. All other parts of the structure can be seen as a specific way of wiring up these regions which act as local Hall elements. (c) Same as (b) but for a positive value of the distance parameter, $d=34\,\si{\micro\meter}$. The other parameters are $R=48\,\si{\micro\meter}$, $r=6\,\si{\micro\meter}$, $a=124\,\si{\micro\meter}$, and $\sigma_0^0=200\,\si{\siemens\per\meter}$.}
\end{figure}
Based on the cofactor calculations for a microstructure, which require knowledge of the local zero magnetic-field conductivity only, one can assign different Hall coefficients to different parts of the microstructure. The choice of the Hall coefficients determines where and how the local values of the cofactors enter into the average, \textit{i.e.}, into the effective Hall coefficient. For example, one can choose the Hall coefficient to be zero in some parts of the structure while taking a certain finite value, $A_{\text{H}}^{0}$, everywhere else. For our interlinked tori geometry, three such assignments are shown in figure\,5. The zero magnetic field conductivity is the same in all parts of the structure. The simplest assignment is probably to choose a constant Hall coefficient for all parts, resulting in the single-constituent porous structure shown in figure\,5(a). From the numerical calculations we obtain $A_{\text{H}}^*=-5.73\, A_{\text{H}}^{0}$. Not only is the sign of the Hall coefficient reversed but its magnitude is also substantially increased, at the cost of increased resistivity of the material. As there are no regions where the cofactor takes considerable positive values, this choice is hard to beat if one aims at maximizing the modulus of the sign-inverted effective Hall coefficient. If the Hall coefficient is nonzero only in the cylinders connecting the tori as shown in figure\,5(b), the effective Hall coefficient will be much smaller, $A_{\text{H}}^*=-1.86\, A_{\text{H}}^{0}$. If however the Hall coefficient is nonzero only in the regions where the cylinders and tori intersect as shown in figure\,5(c), one almost recovers the Hall coefficient of the single-constituent structure, $A_{\text{H}}^*=-5.39\,  A_{\text{H}}^{0}$.

In principle, it might be possible to create a microstructure for which the sign of the effective Hall coefficient is controlled by the assignment of the local Hall coefficient. This would be possible in a hypothetical material exhibiting regions with considerable positive values of the cofactor as well as regions with considerable negative values of the cofactor which are spatially well separated.  
 
\begin{figure}[h]
	\includegraphics[width=16cm]{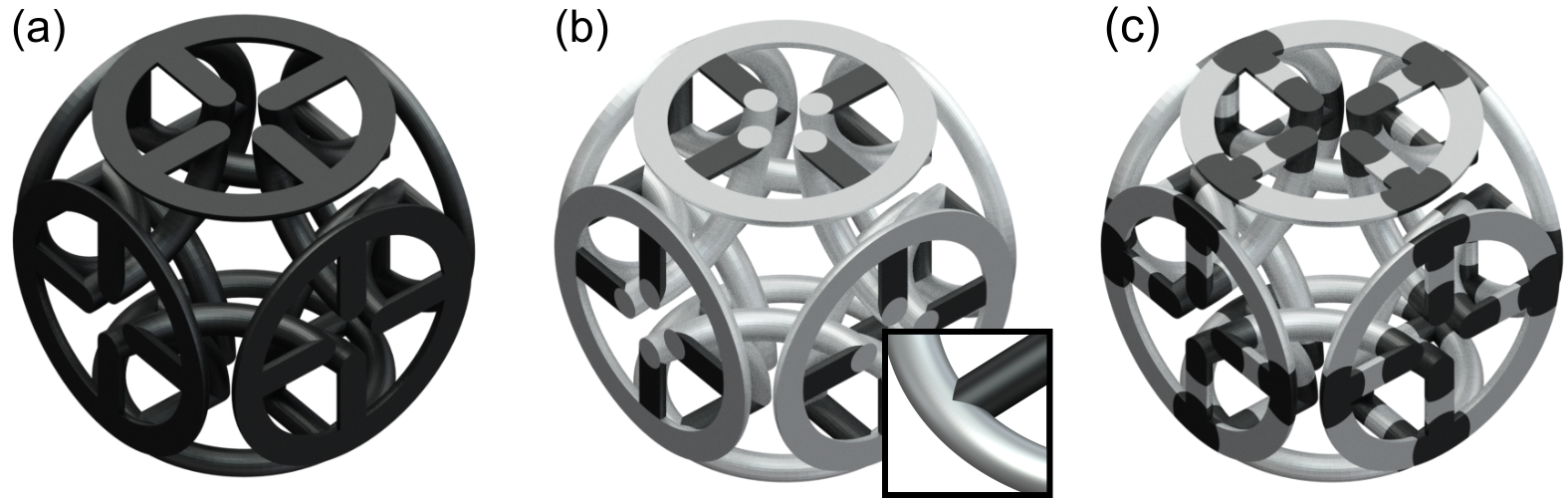}
	\caption{Unit cells of different chainmail-inspired metamaterials based on intertwined tori. Starting from the cofactor matrix field shown in figure\,4, one can assign different Hall coefficients to different parts of the material. Here, two constituent materials are employed, one with a finite Hall coefficient, shown in black, and one with zero Hall coefficient, shown in grey. Both materials have the same zero magnetic field isotropic conductivity. (a) A single-constituent structure. (b) A structure with tori made from the zero Hall coefficient material and cylinders made from the material with a finite Hall coefficient. (c) A structure in which only the regions where the cylinders and tori intersect have nonzero Hall coefficient. As these are the regions that give the main contribution to the effect, the modulus of the effective Hall coefficient is almost as large as in (a).}
\end{figure}

In the experiments, a slightly different structure has been realized \cite{Kern17}. Using three-dimensional laser lithography, an ultra-high resolution 3D printing technique, electrically insulating polymer scaffolds have been fabricated on the micrometer scale. By means of atomic layer deposition, they were coated with an n-type semiconductor resulting in an electrically hollow structure. We have argued previously that this does not affect the qualitative behavior of the structures \cite{Kadic15}, \cite{Kern17}. This is confirmed by the numerical calculations shown in figure\,6. Regarding the trace of the cofactor matrix, we recover, in essence, the results shown in figure\,4. In figure\,6(b), the behavior of the effective Hall coefficient as a function of the distance parameter is shown. Compared to the non-hollow version, the effective Hall coefficient is much larger as the current flow is restricted to a very thin layer. At the same time, the hollow structure has a lower effective conductivity (see also section \ref{Bounds}).

We find that, \textit{via} a fit to further numerical data not depicted here, in the saturated regime, the modulus of the effective Hall coefficient is approximately given by 

\begin{equation*}
\left|A_{\text{H}}^*\right|\approx 0.2\frac{a}{t}\left|A_{\text{H}}^0\right|.
\end{equation*}
Similar estimates can be obtained by describing the structure as a network of voltage sources (representing the regions of large cofactor) and resistances. The corresponding voltages can be estimated by, \textit{e.g.}, approximating these regions as hollow cylinders (see also \cite{Kern17-2}). 

\begin{figure}[h]
	\includegraphics[width=16cm]{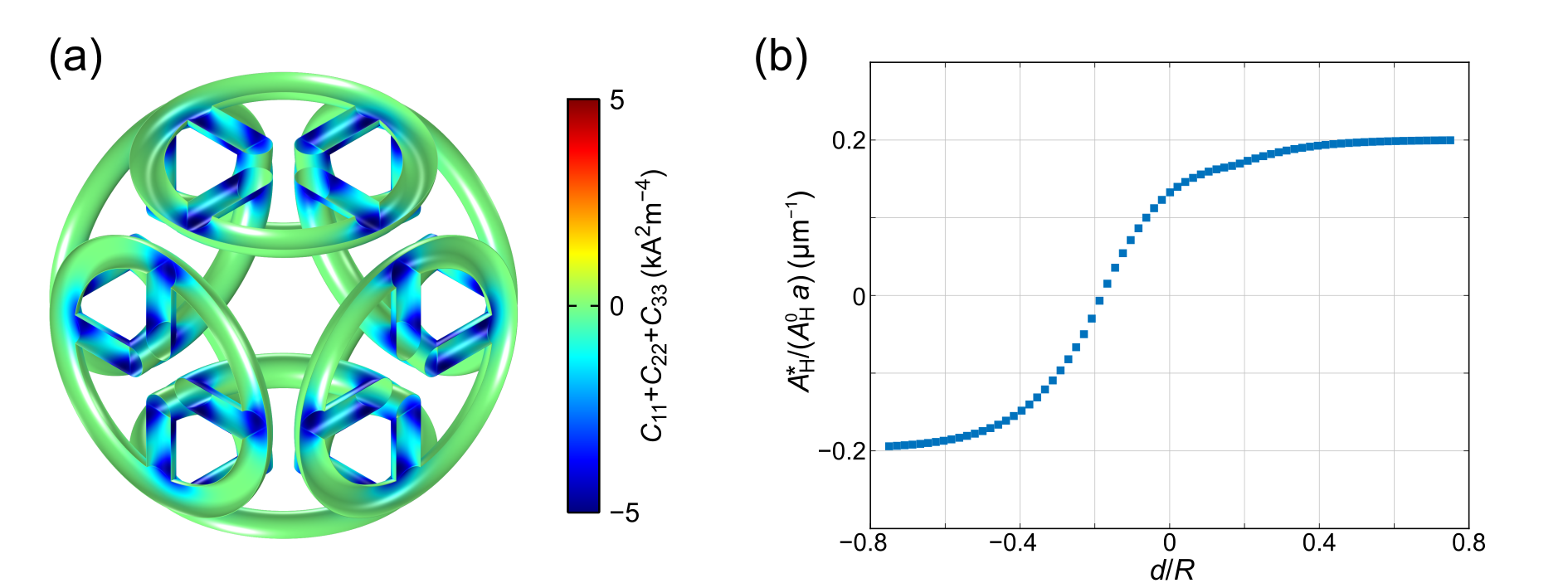}
	\caption{(a) Numerical calculation of the trace of the cofactor matrix of the matrix-valued electric current for a hollow version of the structure shown in figure\,4. This hollow geometry corresponds to the experimental realization \cite{Kern17}. There, electrically insulating polymer scaffolds were coated with a thin layer of an n-type semiconductor. Qualitatively, the hollow and non-hollow structures show the same behavior. However, the hollow structures have a higher absolute effective Hall coefficient at the expense of a lower effective conductivity. The thickness of the shell is $t=185\,\si{\nano\meter}$. All other parameters are as in figure\,4(a). (b) Calculated  effective Hall coefficient in dependence of the distance parameter, $d/R$, for the structure shown in (a) and a shell thickness of $t=1\,\si{\micro\meter}$. These results correspond to those shown in figure\,9 of \cite{Kadic15}, which were obtained for a finite size Hall bar.}
\end{figure}

\section{Symmetry considerations}
When it comes to the Hall effect, one has to be careful with the notions of symmetry and isotropy. Obviously, for a nonzero magnetic field, the conductivity tensor becomes anisotropic. However, a material characterized by an isotropic conductivity tensor and an isotropic Hall matrix, \textit{i.e.}, a scalar Hall coefficient, is certainly isotropic. This difference arises, because in coordinate transformations one can either keep the magnetic field fixed while transforming the electric field and the current field or one can transform all three fields simultaneously. 
%This is somewhat similar similar to the local breaking of time reversal symmetry in structures due to an external magnetic field, which is often employed in two-dimensional topological insulators. Transforming the magnetic field as well would preserve the symmetry. 
Here, we study the restrictions on the Hall matrix---and the zero magnetic field conductivity matrix---imposed by the symmetry of the structure itself. Hence, we consider the Hall matrix in a transformed coordinate system, simultaneously transforming all three fields. 

In short, one could argue that we have already seen that the Hall matrix is a tensor---using that $\bm{\sigma}$ is a tensor and $\bm{b}$ is an axial vector---and hence, the usual symmetry considerations can be applied. In more detail, we can derive the transformation of the Hall matrix from the constitutive equation.
In transforming the fields, one has to keep in mind that $\bm{e}$ and $\bm{j}$ are polar vectors while $\bm{b}$ is an axial vector and hence, $\bm{e'}=\bm{R}\bm{e}$, $\bm{j'}=\bm{R}\bm{j}$, and $\bm{b'}=\det{\left(\bm{R}\right)}\bm{R}\bm{b}$, where $\bm{R}$ is an orthogonal transformation matrix, $\bm{R}^{\intercal}=\bm{R}$.
Starting from 
\begin{equation}
\bm{e} = \left(\bm{\rho}_0+\mathscr{E} \left(\bm{A}_{\text{H}}\bm{b}\right)\right)\bm{j}\text{ and }\bm{e}' = \left(\bm{\rho}_0'+\mathscr{E} \left(\bm{A}_{\text{H}}'\bm{b}'\right)\right)\bm{j}',
\end{equation}
it follows that
\begin{equation}
\bm{R}\bm{e} = \left(\bm{\rho}_0'+\mathscr{E} \left(\bm{A}_{\text{H}}'\det\left(\bm{R}\right)\bm{R}\bm{b}\right)\right)\bm{R}\bm{j}
\end{equation}
and
\begin{equation}
\bm{R}^{\intercal}\bm{\rho}_0'\bm{R}=\bm{\rho}_0\text{ and }\bm{R}^{\intercal}\mathscr{E} \left(\bm{A}_{\text{H}}'\det\left(\bm{R}\right)\bm{R}\bm{b}\right)\bm{R}=\mathscr{E} \left(\bm{A}_{\text{H}}\bm{b}\right).
\end{equation}
Using $\bm{R}^{\intercal}\mathscr{E}\left(\bm{x}\right)\bm{R}=\mathscr{E}\left(\text{Cof}\left(\bm{R}\right)^{\intercal}\right)$ for any $\bm{x}$, $\text{Cof}\left(\bm{R}\right)^{\intercal}=\det\left(\bm{R}\right)\bm{R}^{\intercal}$ and $\left(\det\left(\bm{R}\right)\right)^2=1$ one obtains that
\begin{equation}
\bm{R}^{\intercal}\bm{A}_{\text{H}}'\bm{R}=\bm{A}_{\text{H}}.
\end{equation}
Hence, the Hall matrix has the transformation properties of a rank-2 tensor. Using Neumann's principle, one can identify the symmetry restrictions on the tensor. Typically, one determines the crystallographic point group of the structure. The symmetry restrictions can then be found by various methods, \textit{e.g.}, using representation theory.

Importantly, for any cubic crystallographic point group which is characterized by four threefold rotational axes, the Hall matrix as well as the conductivity matrix are multiples of the identity. Our chainmail-inspired metamaterial is an example of a structure with the highest symmetric crytallographic cubic point group. Later, we will introduce an isotropic structure with the lowest cubic crystallographic point group $32$. This point group has, in addition to the four threefold axes, three twofold axes of rotation.

In a potential experiment, we can think of isotropy as follows. Assume we have a large block of a metamaterial and cut out Hall bars at different orientations. Isotropy means that all these Hall bars have the identical behavior in all Hall measurements. 

\section{``Anti-Hall bars''}
It has been pointed out that part of the unit cell of the chain-mail inspired sign-inversion metamaterial shows some similarity to the previously studied ``anti-Hall bar'' \cite{Mani94, Mani17}. In some sense, a torus can be seen as a three dimensional analogue of a planar Hall bar with a hole. It was demonstrated, that upon moving the current injection as well as the voltage sensing contacts from the outer to the inner perimeter of such a Hall bar, the Hall voltage changes sign \cite{Mani17}. For point-like contacts on the boundary of a Hall bar it is easy to see why. 

\begin{figure}[h!]
	\includegraphics[width=16cm]{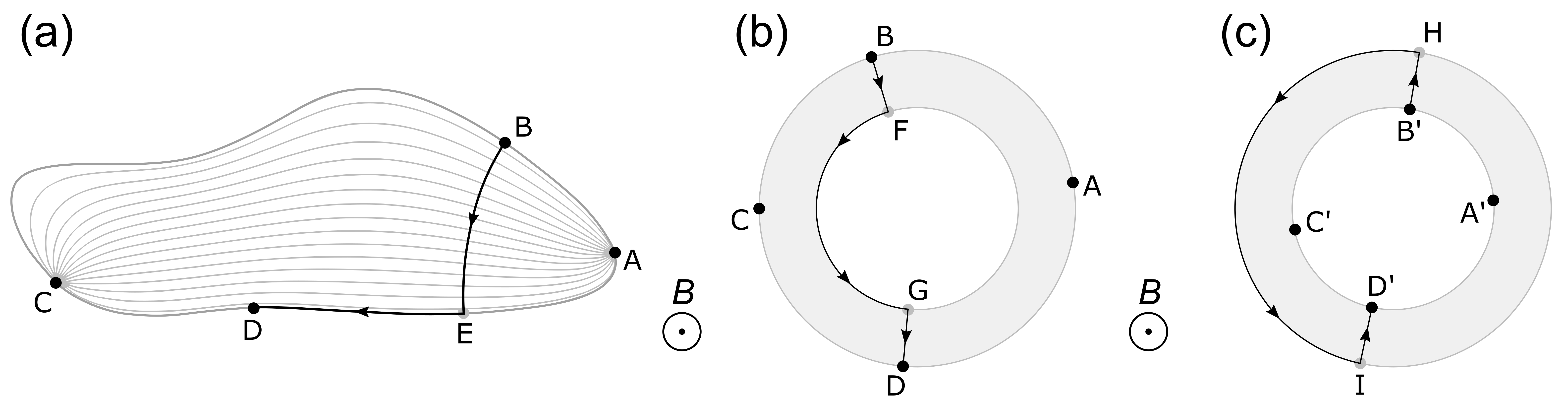}
	\caption{(a) An arbitrarily shaped, simply connected planar Hall bar with point-like contacts on the outer boundary. The current is flowing from A to C and the voltage is measured between B and D. Current streamlines are shown in grey. (b) A doubly-connected Hall bar with point contacts on the outer boundary. As in (a), the current is flowing from A to C and the voltage is measured between B and D. (c) A doubly-connected Hall bar with contacts on the inner boundary. The current is flowing from A' to C' and the voltage is measured between B' and D'. In all three cases, the path of integration is shown. This path is split into parts, each being either along the boundary or perpendicular to the direction of current flow. Panel (a) has been adapted from \cite{Popovic04}.}
\end{figure}

We start by summarizing some results for Hall devices with point contacts on the boundary, following the lines of Popovic \cite{Popovic04}. Consider the Hall bar shown in figure\,7(a). We study the two-dimensional problem with resistivity $\bm{\rho}_0=\text{diag}\left(\rho_0,\rho_0\right)$ and Hall coefficient $A_{\text{H}}$ and a perpendicular magnetic flux density $\bm{b}$. We impose a current $I$, flowing from contact A to contact C and measure the voltage between contacts B and D. This voltage is given by
\begin{equation}
V_{\text{BD}}=\int_{\text{B}}^{\text{D}}\bm{\rho}_0\bm{j}\,\text {d}\bm{l}+\int_{\text{B}}^{\text{D}}A_{\text {H}}\left(\bm{b} \times \bm{j}\right)\text {d}\bm{l}
\end{equation}
The Hall voltage is the difference between $V_{\text{BD}}$ for some finite magnetic field and $V_{\text{BD}}$ for zero magnetic field. As we consider point contact devices, the current distribution is independent of the magnetic field
and hence, we obtain the following expression for the Hall voltage
\begin{equation}
V_{\text {H}}=\int_{\text{B}}^{\text{D}} A_{\text {H}}\left(\bm{b} \times \bm{j}\right)\text {d}\bm{l}.
\end{equation}
In order to evaluate this integral, we choose a specific path, such that we integrate perpendicular to the lines of current flow from B to E and along the boundary from E to D as shown in figure\,7(a). As there is no current flow through the boundary, the second part of the line integral vanishes. Note that this means that the Hall voltage between any two points on the same boundary is zero as long as there is no current injected in between them. For the doubly-connected geometry considered below, it implies that just moving the current or the sensing contacts onto the inner boundary, while keeping the other contacts on the outer boundary, will result in a zero Hall voltage. The first part of the line integral can be evaluated easily,
\begin{equation}
V_{\text {H}}=\int_{\text{E}}^{\text{D}} A_{\text {H}}\left(\bm{b} \times \bm{j}\right)\text {d}\bm{l}=\frac{A_{\text {H}}}{t}Ib,
\end{equation}
where $t$ is the thickness of the Hall bar.

Now, we consider the doubly-connected Hall device shown in figure\,7(b). In a first experiment, the current is flowing from contact A to contact C and we measure the Hall voltage between contact B and contact D. In a second experiment, the current is flowing from A' to D' and we measure the Hall voltage between contacts B' and D'. This second geometry has been termed ``anti-Hall bar'' \cite{Mani94} and it has been shown that the two Hall voltages have opposite sign.

In order to obtain the Hall voltage for the first configuration, we can evaluate the line integral as previously by integrating along the boundary and perpendicular to the direction of current flow as shown in figure\,7(b)
\begin{equation*}
V_{\text {H}}=\int_{\text{B}}^{\text{F}} A_{\text {H}}\left(\bm{b} \times \bm{j}\right)\text {d}\bm{l}+\int_{\text{G}}^{\text{D}} A_{\text {H}}\left(\bm{b} \times \bm{j}\right)\text {d}\bm{l}=\frac{A_{\text {H}}}{t}Ib.
\end{equation*}
The position of the contacts on the boundary is irrelevant, as long as their sequence is not changed.
Analogously, as shown in figure\,7(c), we obtain for the second configuration 
\begin{equation*}
V_{\text {H}}=\int_{\text{B'}}^{\text{H}} A_{\text {H}}\left(\bm{b} \times \bm{j}\right)\text {d}\bm{l}+\int_{\text{I}}^{\text{D'}} A_{\text {H}}\left(\bm{b} \times \bm{j}\right)\text {d}\bm{l}=-\frac{A_{\text {H}}}{t}Ib.
\end{equation*}
Hence, the Hall voltages have opposite sign.

These considerations aim at providing an intuition for the change of sign of a local Hall voltage in the chain-mail inspired metamaterial. However, we emphasize once again that it is a very demanding task to translate the sign-inversion of a local Hall voltage into the change of sign of the effective Hall coefficient which is a material parameter. The previous work by Mani \textit{et al.} \cite{Mani94} was not concerned with effective material parameters at all. 

\section{A second architecture exhibiting a sign reversal of the Hall coefficient}

Perhaps the easiest way to invert the sign of a Hall voltage measured on a Hall bar, or in fact any voltage, is to take the two sensing wires and interchange them. As we will see, this simple idea can be extended into a structure with an inverted effective Hall coefficient. 

\begin{figure}[h!]
	\includegraphics[width=16cm]{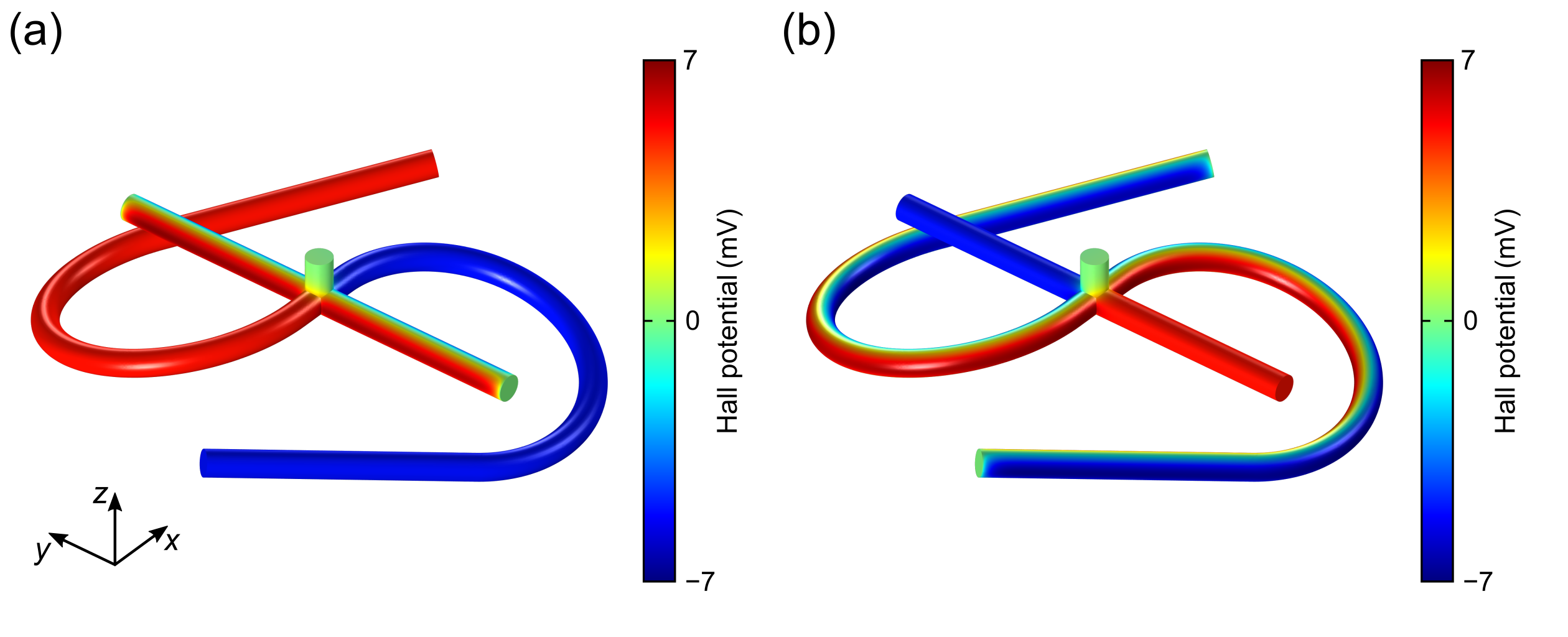}
	\caption{Numerical calculations of the perturbation in the electric potential due to the presence of a magnetic field along the $z$-direction in a simple structure inverting the measured Hall voltage. The geometry is similar to a cross, however, two of the opposing segments are bent, reversing the connection such that the Hall voltage is inverted. (a) Assuming a current flow along the $y$-direction, the pickup of the Hall voltage is reversed. (b) Assuming a current flow along the $x$-direction, the direction of current flow is locally reversed, while the pickup is not. In both cases, the Hall voltage changes sign. Boundary conditions are constant potential for two faces each for imposing the current flow of $I=100\,\si{\micro\ampere}$ and insulating boundaries everywhere else. Parameters are $R_1=30\,\si{\micro\meter}$, $r=3\,\si{\micro\meter}$, $\alpha=15^{\circ}$, $\beta=220^{\circ}$, $A_{\text {H}}=-624\cdot 10^{-6}\,\si{\cubic\meter\per\ampere\per\second}$, and $B=1\,\si{\tesla}$.}
\end{figure}

In figure\,8, numerical calculations of the Hall potential, \textit{i.e.}, the perturbation in the electric potential due to a magnetic field, in a single corresponding element are shown. The magnetic field is pointing in the $z$-direction. Figure\,8(a) shows the result for a current flow in the $y$-direction, through the straight cylinder. A local Hall voltage appears and its pickup is reversed with respect to the $x$-direction \textit{via} the bent segments. In figure\,8(b), the current is flowing in the $x$-direction, through the bent segments. Locally, the direction of current flow is reversed, resulting in an inverted Hall voltage as well. 

\begin{figure}[h!]
	\includegraphics[width=16cm]{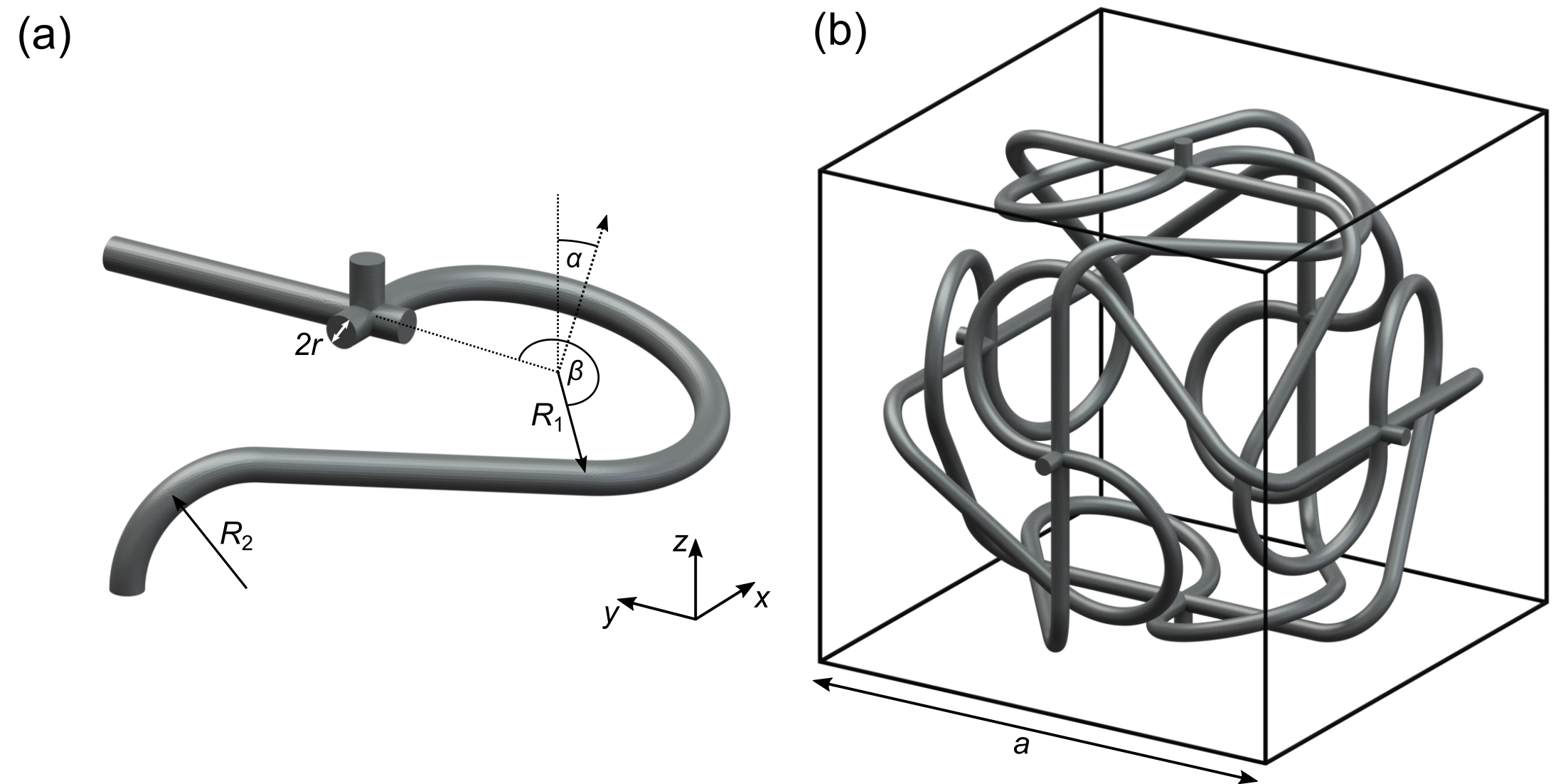}
	\caption{(a) Scheme of the single constitutive element of an electrically isotropic structure showing a sign-inversion of the Hall coefficient. Such elements can be arranged into a three-dimensional structure, a corresponding unit cell of which is shown in (b). The structure has the cubic crystal point group $32$. Therefore, the effective conductivity and the effective Hall matrix are isotropic.}
\end{figure}

One can arrange eight of such elements in the unit cell of a single-constituent, porous metamaterial as shown in figure\,9. This structure has four three-fold rotation symmetry axes and the cubic crystallographic point group $32$. Therefore, the electrical properties, \textit{i.e.}, the effective zero magnetic field conductivity and the Hall matrix, are isotropic. The remaining question is whether its effective Hall coefficient is actually sign-inverted. 

\begin{figure}[h!]
	\includegraphics[width=16cm]{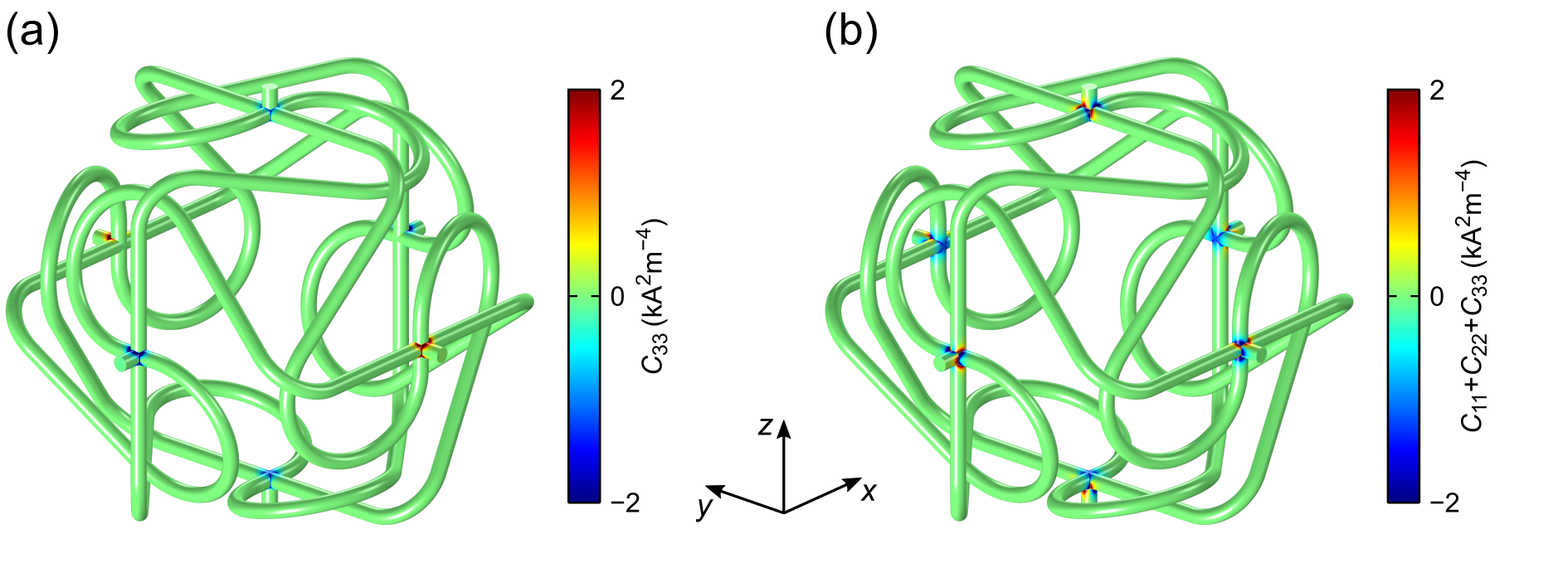}
	\caption{Numerical calculations of (a) one of the three diagonal cofactors, $C_{33}$ corresponding to a magnetic field along the $z$-direction, and (b) the trace of the cofactor matrix of the matrix-valued electric current for the sign-inversion structure shown in figure\,6. For the effect, the Hall coefficient in and around the crosses is crucial. In contrast to the chainmail-inspired geometry, only some parts of the structure contribute to the sign-inversion. Two of the four vertical crosses are counteracting the effect. Looking at (a), one might be tempted to choose a zero Hall coefficient material for the crosses, where the cofactor is positive. However, this would result in a loss of symmetry. In the symmetrized version shown in (b), it becomes clear that the positive and negative regions are slightly different, however, they are certainly not well separated. The volume average of each the three diagonal cofactors takes the same negative value. Parameters are $R_2=20\,\si{\micro\meter}$, $a=170\,\si{\micro\meter}$, and $\sigma_0^0=200\,\si{\siemens\per\meter}$. All other parameters are as in figure\,8.}
\end{figure}

In figure\,10, the results of a numerical calculation of the cofactor are shown. In general, one could, as previously, assign different Hall coefficient to different parts of the structure. Thereby, one would create a multi-constituent structure. If one keeps the same conductivity everywhere, the results shown in figure\,10 do not change, as the Hall coefficient does not enter in the calculation of the cofactor. The Hall coefficient determines the regions where the cofactor enters in the calculation of the effective Hall coeffcient. 
However, the overall average of the cofactor is negative, meaning that it is sufficient to choose the Hall coefficient to be the same everywhere in order to obtain a sign-inversion. From the numerical calculations, we obtain $A_{\text {H}}^*=-3.43A_{\text {H}}^0$ and $\sigma_0^*=1.11\cdot 10^{-3}\sigma_0^0$, where $A_{\text {H}}^0$ is the Hall coefficient and $\sigma_{0}^0$ is the conductivity of the constituent material, for the set of parameters given in figure\,10. The most important parts are again the crossings, where the cofactor has a large modulus, while all other parts may be seen as a clever way of connecting these elements. Interestingly, however, only some of the crosses contribute to the sign-inversion while in others, the cofactor is positive. This can be understood from a calculation of the Hall potential in a finite Hall bar as shown in figure\,11. There, the current is flowing along the $x$-direction and the magnetic field is along $\hat{\bm{z}}$. The vertical structures at the top and the bottom of the unit cells always contribute to the effect. As shown in figure\,11, the voltage pickup or the local direction of current flow, depending on the average direction of current flow, is inverted. Note that for arbitrary average current flow in the $xy$-plane, these two effects mix. However, only two of the four vertical crosses contribute as only in two of those, the pickup or the local current flow---again depending on the direction of current flow---is reversed. 

\begin{figure}[h!]
	\includegraphics[width=16cm]{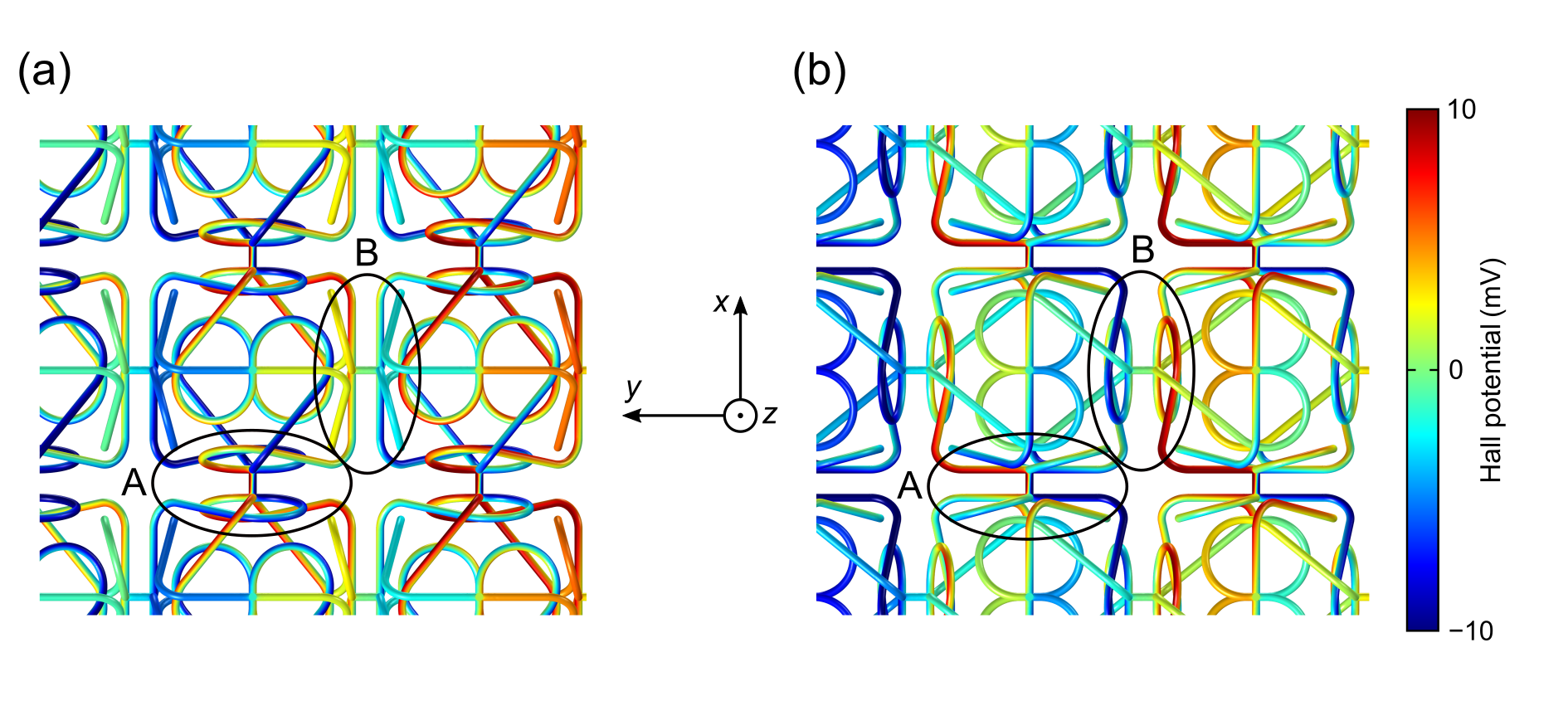}
	\caption{Numerical calculations of the Hall potential in a Hall bar made from $9\times 4 \times 1$ unit cells. Only a few unit cells in the center of the Hall bar are shown. The magnetic field is pointing in the $z$-direction, the current is flowing in the $x$-direction. (a) Calculation for a unit cell as shown in figure\,9. (b) The same calculation, but for a unit cell rotated by $\pi/2$ about the $z$-axis. In both cases, the horizontal crosses contribute to the sign-inversion. Things are more delicate for the vertical crosses, indicated by A and B. In (a), the pickup of the Hall voltage in region A is inverted, thereby contributing to the effect. However, in region B, the current is flowing along $\bm{\hat{x}}$ and the pickup is not inverted. This is the reason why only two of the vertical crosses contribute to the effect, as it is evident from figure\,10. In (b), pickup and current flow are not inverted in region A. In region B, the direction of current flow is locally reversed while the pickup is not. 
	Boundary conditions are fixed potentials for imposing the current flow and insulating boundaries everywhere else. Parameters are $I=1\,\si{\milli\ampere}$, $A_{\text {H}}=-624\cdot 10^{-6}\,\si{\cubic\meter\per\ampere\per\second}$, and $B=1\,\si{\tesla}$. The other parameters are as in figure\,10.}
\end{figure}

\section{Off-diagonal terms and anti-symmetric Hall matrices}

So far we have considered isotropic Hall matrices. In this case, the Hall electric field $\bm{e}_{\text{H}}$ is perpendicular to the magnetic field. Off-diagonal elements of the Hall matrix however, can lead to components of the Hall electric field collinear with $\bm{b}$. In all cases, the Hall electric field will be perpendicular to the direction of current flow as it can be seen from the cross product in equation\,(\ref{constcr}).

Consider, for example, a material with an isotropic zero magnetic field conductivity and the following Hall matrix
\begin{equation*}
\bm{A}_{\text{H}}=\left(\begin{array}{ccc} 0 &  0 & 0\\  0  & 0 & A_{23}\\ 0 & 0 & 0\\ \end{array}\right).
\end{equation*}
Assuming a current flow along the $x$-direction, and a magnetic field along the $z$-direction, we obtain the following expression for the Hall electric field
\begin{equation*}
 \bm{e}_{\text{H}}=-A_{23} j_x b_z \hat{\bm{z}}.
\end{equation*}
Here, $\bm{e}_{\text{H}}$ is parallel to the magnetic field. This is the so-called parallel Hall effect, which has been predicted for effective materials theoretically and numerically \cite{Briane10, Kern15} and later demonstrated experimentally \cite{Kern17-2}. If we add another off-diagonal component to the Hall matrix, such that it is antisymmetric, we obtain the parallel Hall effect independently of the orientation of the magnetic field in the plane perpendicular to the direction of current flow.
Consider again a material with isotropic zero magnetic field conductivity and the following Hall matrix
\begin{equation}
\bm{A}_{\text{H}}=\left(\begin{array}{ccc} 0 &  0 & 0\\  0  & 0 & A_{23}\\ 0 & -A_{23} & 0\\ \end{array}\right).
\label{antisymmhm}
\end{equation}
Again we assume that a current is flowing along the $x$-direction and restrict the magnetic field to the $yz$-plane. Then, we obtain the following expression for the Hall electric field
\begin{equation*}
	\bm{e}_{\text{H}}=-A_{23} j_x \left(b_y \hat{\bm{y}} + b_z  \hat{\bm{z}} \right).
\end{equation*}
Clearly, $\bm{e}_{\text{H}}$ and $\bm{b}$ are collinear independently of the orientation of the magnetic field in the $yz$-plane. One may ask, whether such effective properties are realizable in metamaterials with isotropic constituents. This question was answered by Briane and Milton. They considered the three-constituent structure with the unit cell schematically shown in figure\,12. It consists of a square cylinder oriented along the $x$-direction with conductivity $\bm{\sigma}=\bm{I}$, a spirally shaped pick-up structure with a high conductivity in the $yz$-plane, $\bm{\sigma}^{\kappa}=\text{diag}\left(\kappa,\kappa,1\right)$, and some surrounding material with the same conductivity as the square cylinder. The Hall coefficient is nonzero in the square cylinder and zero everywhere else. A specific choice is made for the Hall coefficient in the square cylinder such that the effective Hall matrix will be normalized. Note that we can replace the anisotropic material of the spirally shaped part by a laminate. One obtains a four-constituent material with isotropic constituents. In the limit of an infinitely high in-plane conductivity $\left(\kappa \rightarrow \infty \right)$, this structure has the antisymmetric effective Hall matrix given in equation\,(\ref{antisymmhm}) with $A_{23}=1$. One can understand this behavior intuitively. Consider a current flowing along the $x$-direction. The part of the current flowing through the square cylinder will, in the presence of a magnetic field in the $yz$-plane, lead to a local Hall voltage. On the sides of the cylinder, the electric potential is picked up and guided by the spirally shaped element such that a potential gradient $\left(-\bm{e}_{\text{H}}\right)$ collinear with the magnetic field appears. Clearly, if we invert the direction of current flow, the Hall electric field will flip.
\begin{figure}[h!]
	\includegraphics[width=8cm]{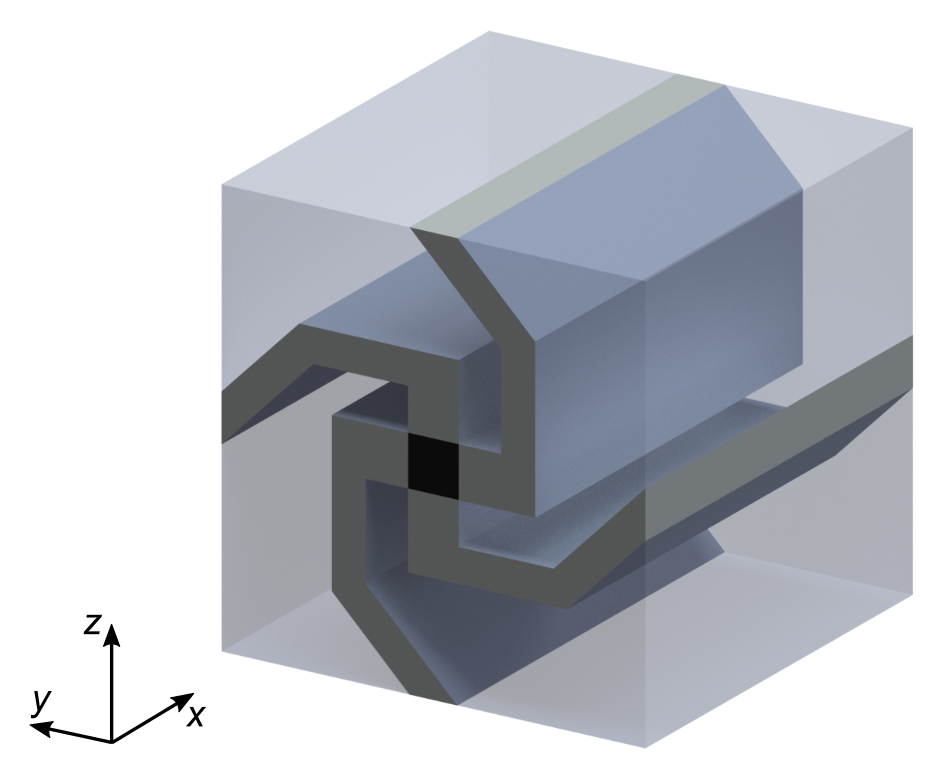}
	\caption{A single unit cell of the three-constituent metamaterial introduced in \cite{Briane10}, (asymptotically) exhibiting an antisymmetric Hall matrix. Only the central square cylinder, shown in black, has a nonzero Hall coefficient. The spiral-shaped part, shown in grey, is highly conducting in the $yz$-plane. Otherwise, the conductivity is the same everywhere.}
\end{figure}

We can approximate such a behavior in a single-constituent porous metamaterial. Consider the structure with the unit cell shown in figure\,13(a). It is made from a single constituent with isotropic zero magnetic field conductivity and an isotropic Hall matrix. If we again consider a macroscopic current flow along the $x$-direction, the current will flow through the cylinder oriented along the $x$-direction and again, a local Hall voltage appears. As previously, the electric potential is guided \textit{via} the spirally shaped part such that the effective electric field is parallel to $\bm{b}$. In the pick-up structure, especially far away from the central cylinder, there is almost no current flow, which makes it possible to use a single constituent, \textit{i.e.}, to choose the Hall coefficient to be the same everywhere. If we made this structure translationally invariant along the $x$-direction, the net effect would be zero.

\begin{figure}[h!]
	\includegraphics[width=15cm]{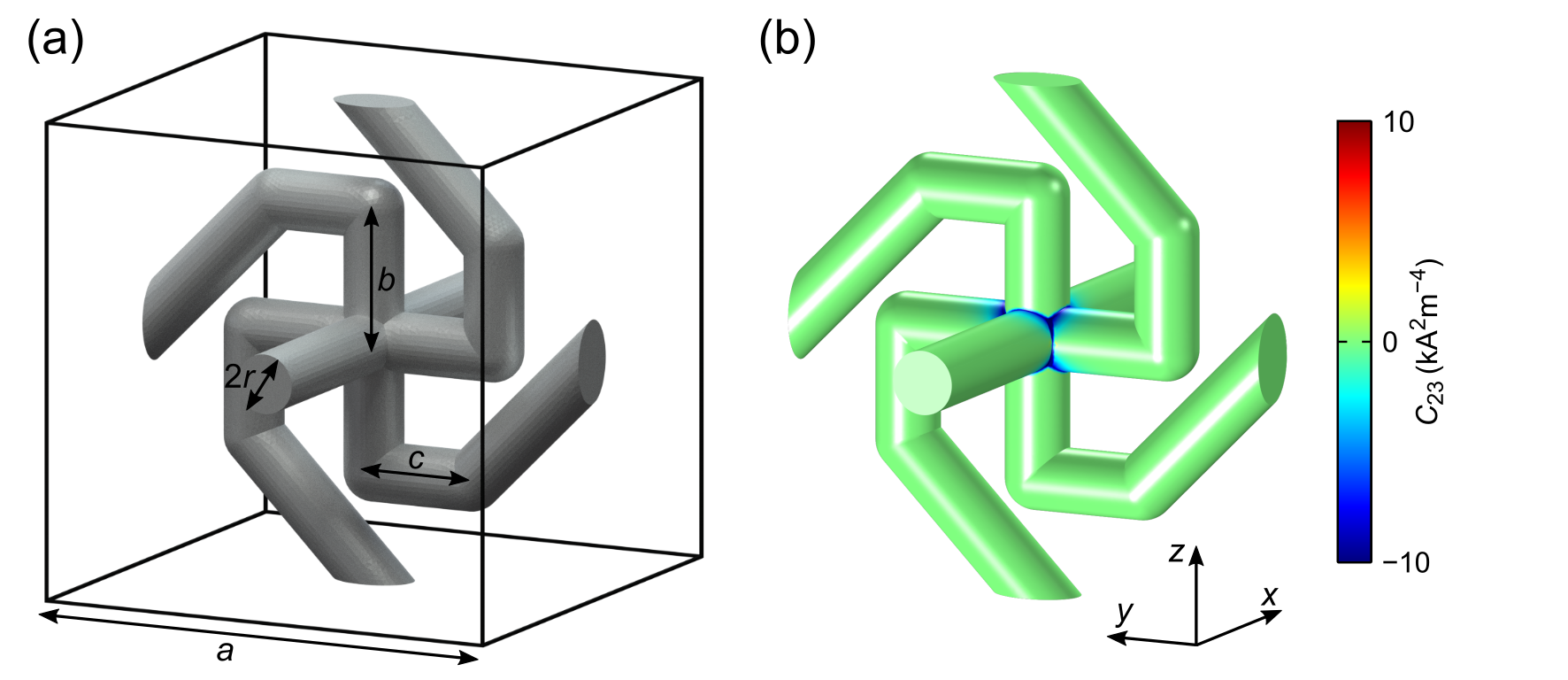}
	\caption{(a) A single unit cell of a single-constituent metamaterial inspired by \cite{Briane10}. Assuming a current flow along the $x$-direction, the spirally shaped part, made from two S-shaped elements, guides the Hall voltage such that the corresponding effective Hall electric field is collinear with the external magnetic field in the $yz$-plane. (b) Numerical calculation of the cofactor $C_{23}$ of the matrix-valued electric current for the structure shown in (a). The cofactor takes large values only in the region where the pickup structure intersects the cylinder. The off-diagonal elements of the corresponding effective Hall matrix are, by symmetry, antisymmetric.  Two of the three diagonal cofactors, $C_{22}$ and $C_{33}$ are small. Due to the intersection of the S-shaped elements however, $C_{11}$ is large. Parameters are $r=2.5\,\si{\micro\meter}$, $a=40\,\si{\micro\meter}$, $b=11\,\si{\micro\meter}$, and $c=9\,\si{\micro\meter}$.}
\end{figure}

A calculation of the cofactor $C_{23}$, corresponding to the effective Hall matrix component $\left(\bm{A}_{\text{H}}^*\right)_{32}$ \textit{via} $\left(\bm{A}_{\text{H}}^*\right)_{32}=\left(\sigma_{11}^* \sigma_{22}^*\right)^{-1}\left\langle C_{23}\right\rangle A_{\text{H}}$, is shown in figure\,13(b). Clearly, the average does not vanish, implying that, assuming appropriate orientations of the current flow and the magnetic field, the Hall electric field will have a component collinear with the magnetic field. Once again, the cofactor has a large modulus only in the central cylinder where the voltage is picked up.  The structure has a fourfold rotational symmetry around the $x$-axis and a perpendicular mirror symmetry. It has the tetragonal crystallographic point group $4/m$. This symmetry imposes the following form on the effective Hall matrix
\begin{equation}
	\bm{A}_{\text{H}}^*=\left(\begin{array}{ccc} A_{11} &  0 & 0\\  0  & A_{22} & A_{23}\\ 0 & -A_{23} & A_{22}\\ \end{array}\right).
	\label{tetragonal}
\end{equation}

\begin{figure}[h!]
	\includegraphics[width=16cm]{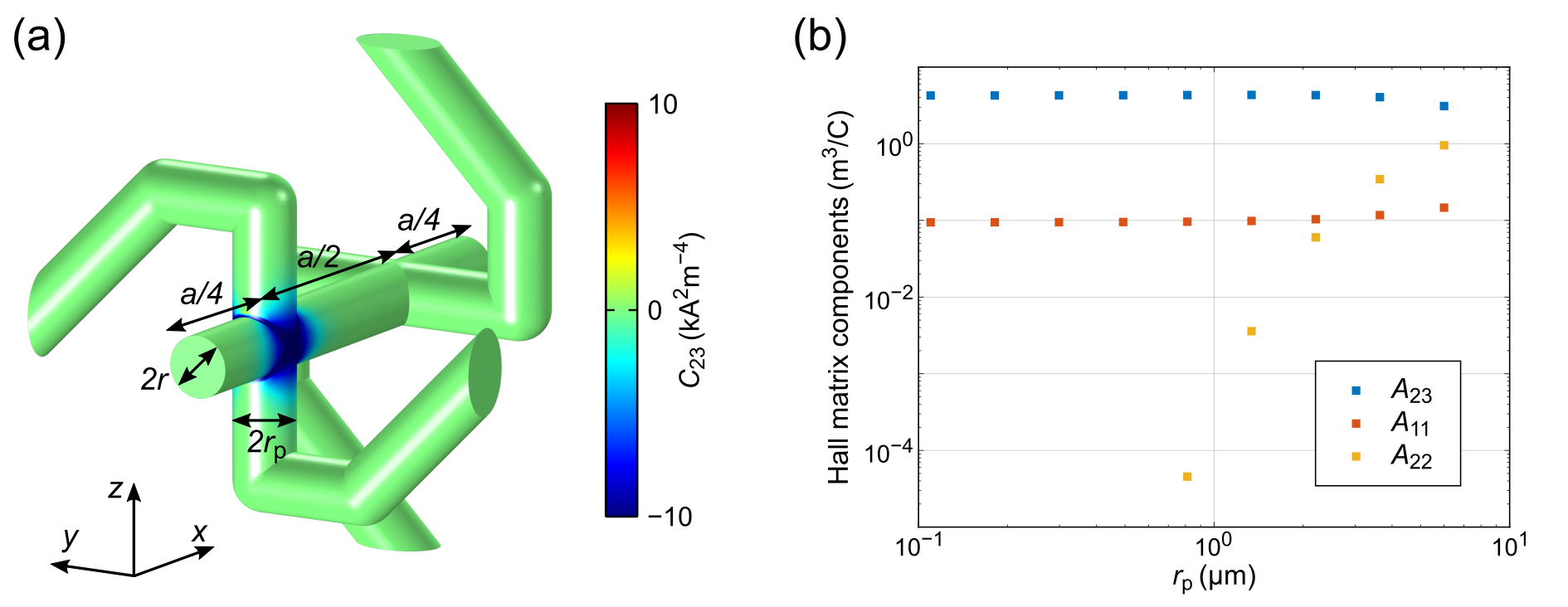}
	\caption{(a) Numerical calculation of the cofactor $C_{23}$ of the matrix-valued electric current for a similar structure as in figure\,13. The two S-shaped elements intersect the cylinder at $x=a/4$ and $x=3a/4$ respectively.	Again, the off-diagonal elements of the corresponding effective Hall matrices are, by symmetry, antisymmetric.  Due to the separation of the pickup structures in the $x$-direction, $C_{11}$ will be much smaller compared to the previous structure. Parameters are as in figure\,13 and $r_{\text{p}}=r$. (b) Behavior of the elements of the Hall matrix in dependence of the radius of the cylinders and spheres in the pickup structure, $r_{\text{p}}$, for $r=6\,\si{\micro\meter}$. As $r_{\text{p}}$ gets smaller, two of the three diagonal elements of the Hall matrix, $A_{22}$ and $A_{33}$, which are identical by symmetry, vanish. However, $A_{11}$ converges to a finite value.}
\end{figure}

From our numerical calculations for the same choice of parameters as in figure\,13, we obtain the following effective Hall matrix
\begin{equation*}
	\bm{A}_{\text{H}}^*=\left(\begin{array}{ccc} 6.85 &  0 & 0\\  0  & 0.04 & 6.84\\ 0 & -6.84 & 0.04\\ \end{array}\right)A_{\text{H}}^0.
\end{equation*}

In order to obtain an antisymmetric Hall matrix, one would like the diagonal elements to become very small and ideally to vanish. The spiral-shaped pick-up structure of the metamaterial shown in figure\,14 consists of two S-shaped structures, one corresponding to the effect along the $y$-direction, the other one corresponding to the effect along the $z$-direction. For the structure shown in figure 9, both of these components pick up the Hall voltage in the same region of the central cylinder. This will lead to a large value of $C_{11}$, as a current flowing through one of the S-shaped elements will generate a local Hall voltage which will be picked up by the other element. In order to minimize this effect, we can separate the two elements along the $x$-direction as shown in the calculation of the cofactor $C_{23}$ in figure 10. Here, the S-shaped elements intersect the central cylinder at $x=a/4$ and $x=3a/4$ respectively, thereby avoiding this coupling. The corresponding cofactors are significantly larger than zero only in small, well separated regions. Once again, we can think of the structure as wired-up local Hall elements. The space group of the structure contains a screw axis and hence, is non-symmorphic. However, the structure has the same tetragonal crystallographic point group as the previous structure, $4/m$. The numerical calculations yield, again for the same parameters as in figure\,13,
\begin{equation*}
	\bm{A}_{\text{H}}^*=\left(\begin{array}{ccc} 0 &  0 & 0\\  0  & 0.05 & 8.81\\ 0 & -8.81 & 0.05\\ \end{array}\right)A_{\text{H}}^0.
\end{equation*}
We can decrease the radius of the elements of the pick-up structure, $r_{\text{p}}$, while keeping the radius of the central cylinder $r$ fixed. The behavior of the components of the effective Hall matrix in dependence of, $r_{\text{p}}$, is shown in figure\,14(b). Two of the three diagonal elements, $A_{22}$ and $A_{33}$, which are identical by symmetry, vanish in the limit of thin pick-up wires. However, $A_{11}$ converges to a finite value.

%Parallel Hall APL structure
In a similar structure, which has been considered previously \cite{Kern15, Kern17-2}, one can go one step further and tailor the angle between the Hall electric field and $\bm{b}$ in the $yz$-plane. The results of numerical calculations of the cofactors $C_{32}$ and $C_{33}$ for this structure are shown in figure\,15. We refer to the original publication for the definition of the parameters \cite{Kern15}. We assume a current flow along the $x$-direction and a magnetic field along the $z$-direction. Again, a current flowing through the central cylinder along the $x$-direction will lead to a local Hall voltage. The electric potential is guided by the two pick-up structures.  By adjusting the corresponding parameters $d_y$ and $d_z$, one can tailor the $z$- and $y$-component of the effective Hall electric field. This means that we can obtain an arbitrary orientation of the effective Hall electric field in the $yz$-plane, as we have shown previously \cite{Kern15}. The structure has a two-fold rotational symmetry about the $x$-axis. It has the monoclinic crystal point group $2/m$. This symmetry implies the form
\begin{equation*}
	\bm{A}_{\text{H}}^*=\left(\begin{array}{ccc} A_{11} &  0 & 0\\  0  & A_{22} & A_{23}\\ 0 & A_{32} & A_{33}\\ \end{array}\right).
\end{equation*}

\begin{figure}[h!]
	\includegraphics[width=14cm]{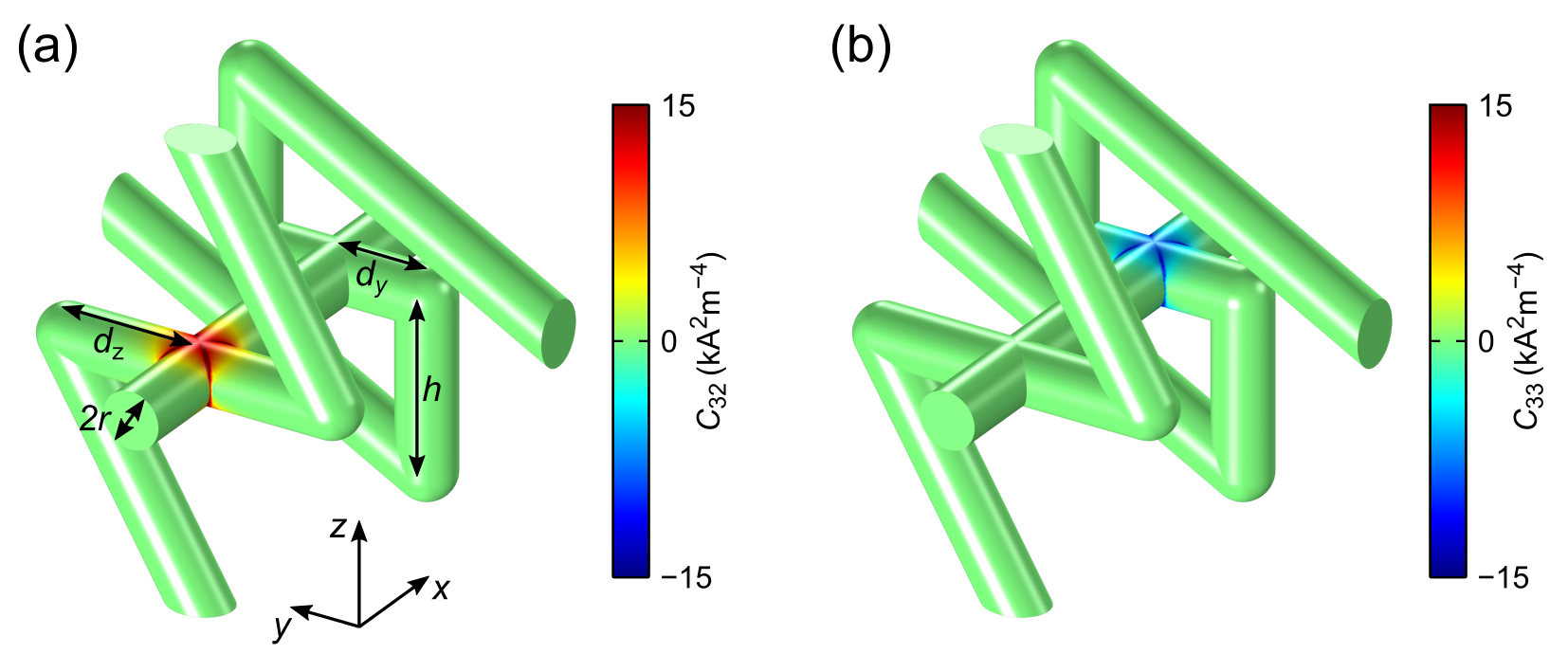}
	\caption{Numerical calculation of two cofactors of the matrix-valued electric current for the parallel Hall effect structure introduced in \cite{Kern15}, (a) $C_{32}$ and (b) $C_{33}$. Assuming a current flow along the $x$-direction and a magnetic field along the $z$-direction, $C_{32}$ corresponds to a Hall electric field in the $z$-direction, \textit{i.e.}, parallel to the magnetic field, whereas $C_{33}$ corresponds to a Hall electric field in the $y$-direction, \textit{i.e.}, orthogonal to the magnetic field. By adjusting the corresponding geometry parameters ($d_y$ and $d_z$), one can arbitrarily tailor the direction of the Hall electric field in the $yz$-plane. Parameters are defined as in \cite{Kern15}, $d_z=-12\,\si{\micro\meter}$, $d_y=-8\,\si{\micro\meter}$, $r=2.5\,\si{\micro\meter}$, $h=15\,\si{\micro\meter}$. The lattice constant is $a=40\,\si{\micro\meter}$.}
\end{figure}

From our numerical calculations, we obtain the following effective Hall matrix for the parameters given in figure\,15
\begin{equation*}
	\bm{A}_{\text{H}}^*=\left(\begin{array}{ccc} 0 &  0 & 0\\  0  & 0.08 & 8.80\\ 0 & 0.31 & -8.73\\ \end{array}\right)A_{\text{H}}^0.
\end{equation*}

For $\left\langle \bm{j}\right\rangle=\left(j_x,~0,~0\right)$, $\bm{b}=\left(0,~0,~b_z\right)^{\intercal}$, and $A_{\text{H}}^0 < 0$, the corresponding angle between the effective Hall electric field and the magnetic field is given by
\begin{equation*}
	\arctan{\left|\frac{A_{33}}{A_{23}}\right|}\approx 45.2^{\circ}.
\end{equation*}

\section{Bounds on the effective parameters} \label{Bounds}
Previous work on Hall-effect metamaterials has addressed the question of bounds \cite{Briane10-2}. The authors showed that the effective Hall matrix is bounded, 
\begin{equation*}
\left|\bm{A}_{\text{H}}^* \right| \leq 18 \frac{\beta_{\text{H}}}{\alpha}a_{\text{H}},
\end{equation*}
where $\alpha$ is a lower bound to the microscopic conductivity, $\beta_{\text{H}}$ is an upper bound to the microscopic conductivity in the regions of nonzero Hall coefficient, $a_{\text{H}}$ is an upper bound to the microscopic Hall coefficient and $\left|\cdot\right|$ is the matrix norm induced by the Euclidian norm.

Therefore, in isotropic structures, the effective Hall coefficient is, up to a constant, bounded by the contrast of conductivities. This bound is in some sense optimal since $A_{\text{H}}^*$ can become infinitely large if the conductivity of one of the constituents tends to zero. Alternatively, one constituent conductivity can tend to infinity, while the others stay finite or tend to zero. Corresponding multiscale laminates have been studied \cite{Briane10-2}.

The first example of a structure allowing for arbitrary values of the effective Hall coefficient, both positive and negative, is the hollow version of the chainmail-like metamaterial introduced in \cite{Kadic15} and discussed above. As one of the constituents is vacuum, and hence, insulating, the contrast of conductivities is infinitely large and the effective Hall coefficient becomes unbounded. Sign-reversal is achieved by adjusting the distance parameter, $d$. By reducing the thickness of the shell, $t$, the effective Hall coefficient can be made arbitrarily large. At the same time, the effective conductivity is reduced. This behavior is fundamental. It follows directly from the work of Briane and Milton that the product of the effective Hall coefficient and the effective conductivity, {\it i.e.}, the effective Hall mobility, is, up to a factor of two, bounded by the maximum of the local Hall mobility of the constituents \footnote{Bounding the product $\left|\Sigma^{\varepsilon} \right|\left|r_{\varepsilon} \right|$ instead of bounding $\left|\Sigma^{\varepsilon} \right|$ and $\left|r_{\varepsilon} \right|$ separately in equation\,(2.24) of \cite{Briane10-2}, immediately yields the desired bound.} \cite{Briane10-2}. In the anisotropic case, the situation is more complex.

This bound limits the applicability of Hall effect metamaterials for Hall sensors. In general, the Hall mobility, which appears in the expressions for the sensitivity and signal-to-noise ratio (depending on the type of noise limiting the measurements), should be as large as possible. However, if we are limited to tailoring the local conductivity and Hall coefficient, we cannot outperform the constituent materials in this sense (at least not by more than factor of two).

\section{Magnetic permeability distributions}
The homogenization formula for the effective Hall matrix, equation\,(\ref{effHallmat}), can be readily extended to account for arbitrary distributions of the magnetic permeability within the unit cell. For the magnetic field, we have that
\begin{equation}
\nabla \cdot \bm{b} = 0,~\nabla \times \bm{h} = 0,~\text{with }\bm{b}=\mu_0\bm{\mu}\bm{h},
\label{constmag}
\end{equation}
where $\bm{h}$ is the magnetic field, $\mu_0$ is the vacuum permeability and $\bm{\mu}$ is the relative magnetic permeability tensor. These equations have the very same form as the static electric conductivity problem, see equation\,(\ref{elcond}). 
Analogous to the electric potential, one can introduce the scalar magnetic potential $\phi_{\text{m}}$, 
\begin{equation*}
\bm{h}=-\nabla \phi_{\text{m}}.
\end{equation*}
%\noindent
Again, we have to distinguish between microscopic and macroscopic fields. The microscopic magnetic field $\bm{h}$ and the corresponding macroscopic field $\left\langle\bm{h}\right\rangle$ are connected \textit{via}
\begin{equation*}
	-\left(\nabla \bm{\Phi}_{\text {m}}\right)\left\langle \bm{h}\right\rangle = \bm{h}.
\end{equation*}
In analogy to the electric problem, $\bm{\Phi}_{\text{m}}$ is the vector-valued ``scalar'' magnetic potential which solves
\begin{equation*}
\nabla \cdot \left(\bm{\mu}\nabla \bm{\Phi}_{\text{m}}\right)=0
\end{equation*}
and is subject to the boundary condition that $\bm{\Phi}_{\text{m}}\left(\bm{y}\right)+\bm{y}$ is invariant with respect to translations by integer multiples of one unit cell.
The corresponding magnetic field $\bm{H}=-\left(\nabla\bm{\Phi}_{\text{m}}\right)$ and the corresponding magnetic flux density $\bm{B}=-\mu_0\bm{\mu}\left(\nabla\bm{\Phi}_{\text{m}}\right)$ are matrix-valued.\\[0.2cm]
\noindent
The constitutive equation for the macroscopic fields reads as $\left\langle \bm{b}\right\rangle=\mu_0\bm{\mu}^*\left\langle \bm{h}\right\rangle$.
Therefore, analogously to equation\,(\ref{constitmacr}), we obtain for the effective permeability
\begin{equation}
	\bm{\mu}^* = \bm{\mu}^*\left\langle \bm{H}\right\rangle =\frac{1}{\mu_0}\left\langle \bm{B}\right\rangle.
	\label{magpermeff}
\end{equation}
\medskip

\noindent
In order to determine the effective Hall matrix, we use equation\,(10), and replace the previously microscopically-constant magnetic field, by $\bm{b}=-\mu_0\bm{\mu}\left(\nabla\bm{\Phi}_{\text{m}}\right)\left\langle\bm{h} \right\rangle$,
\begin{equation*}
	\mathscr{E}\left(\bm{S}\left\langle \bm{b}\right\rangle\right)^* = \left\langle \mathscr{E}\left(\text{Cof}\left(\nabla \bm{U}\right)^{\intercal}\bm{S}\bm{b}\right)\right\rangle,
	\end{equation*}
\begin{equation*}
	\mathscr{E}\left(\bm{S}\mu_0\bm{\mu}^*\left\langle \bm{h}\right\rangle\right)^*  = \left\langle \mathscr{E}\left(\text{Cof}\left(\nabla \bm{\Phi}\right)^{\intercal}\bm{S}\mu_0\bm{\mu}\left(\nabla \bm{\Phi}_{\text{m}}\right)\left\langle \bm{h}\right\rangle\right)\right\rangle.
\end{equation*}
Eventually, we arrive at the following formula for the effective Hall matrix
\begin{equation}
	\text{Cof}\left(\bm{\sigma}_0^*\right)\bm{A}_{\text{H}}^*\bm{\mu}^* = \left\langle \text{Cof}\left(\bm{\sigma}_0\nabla \bm{\Phi}\right)^{\intercal}\bm{A}_{\text{H}}\bm{\mu}\left(\nabla \bm{\Phi}_{\text{m}}\right)\right\rangle.
	\label{effHallmatMag}
\end{equation}
With this equation, we are prepared to calculate the effective properties of metamaterials composed not only of spatial conductivity and Hall-coefficient distributions (as extensively discussed above), but also simultaneously of spatial distributions of magnetic materials. 

\section{A third architecture exhibiting a sign reversal of the Hall coefficient}

\begin{figure}[h!]
	\includegraphics[width=16cm]{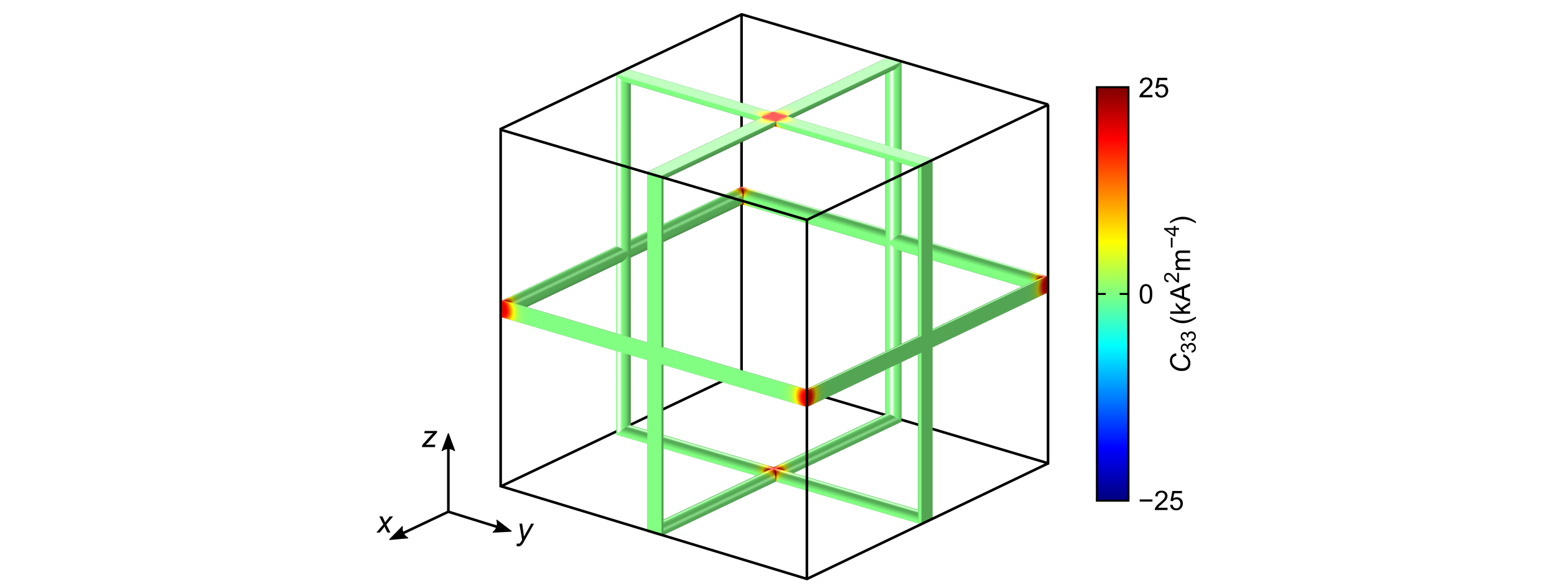}
	\caption{Numerical calculation of the cofactor $C_{33}$ for a body-centered cubic arrangement of crossing electrically-conductive cylinders (cut at the boundaries of the unit cell). They form one component of a structure in which the sign-inversion of the Hall coefficient is due to a second component made from a material with a large magnetic permeability. The crossing regions have by far the dominant contribution to the overall Hall effect. The straight cylinder segments are nearly negligible. Note that the crosses only contribute if there is a perpendicular component of the magnetic field. Only the crosses in the $xy$-plane (and all parallel planes) exhibit a large value of the cofactor $C_{33}$. Hence, it is sufficient to invert the magnetic field for a single direction for each cross. Such a structure is shown in figure\,17(a). Parameters are $r_1=2\,\si{\micro\meter}$, $a=86\,\si{\micro\meter}$, and $\sigma_0^0=200\,\si{\siemens\per\meter}$.}
\end{figure}

In the following, we use equation\,(\ref{effHallmatMag}) to show that it is possible to design an electrically isotropic microstructure with an effective Hall coefficient that is sign-inverted due to an appropriate choice of the magnetic permeability distribution. 

In figure\,16(a), a numerical calculation of the cofactor $C_{33}$ of the matrix-valued current field for the electrically conducting part of such a structure is shown. This structure can be seen as a body-centered cubic arrangement of crosses, each cross consisting of two intersecting cylinders with radius $r_1$. It is made from an electrically conducting material with conductivity $\sigma_{0}^{0}$, Hall coefficient $A_{\text{H}}^0$, and magnetic permeability $\mu^{0} = 1$. It has the highest symmetric crystallographic cubic point group, $\frac{4}{\text{m}}\bar{3}\frac{2}{\text{m}}$, implying that the effective properties are isotropic. According to equation\,(\ref{effHallmat}), the effective Hall coefficient is determined by the volume average of any one of the diagonal cofactors. As $C_{33}$ is positive everywhere, the effective Hall coefficient will \textit{not} be sign-inverted.

\begin{figure}[h!]
	\includegraphics[width=16cm]{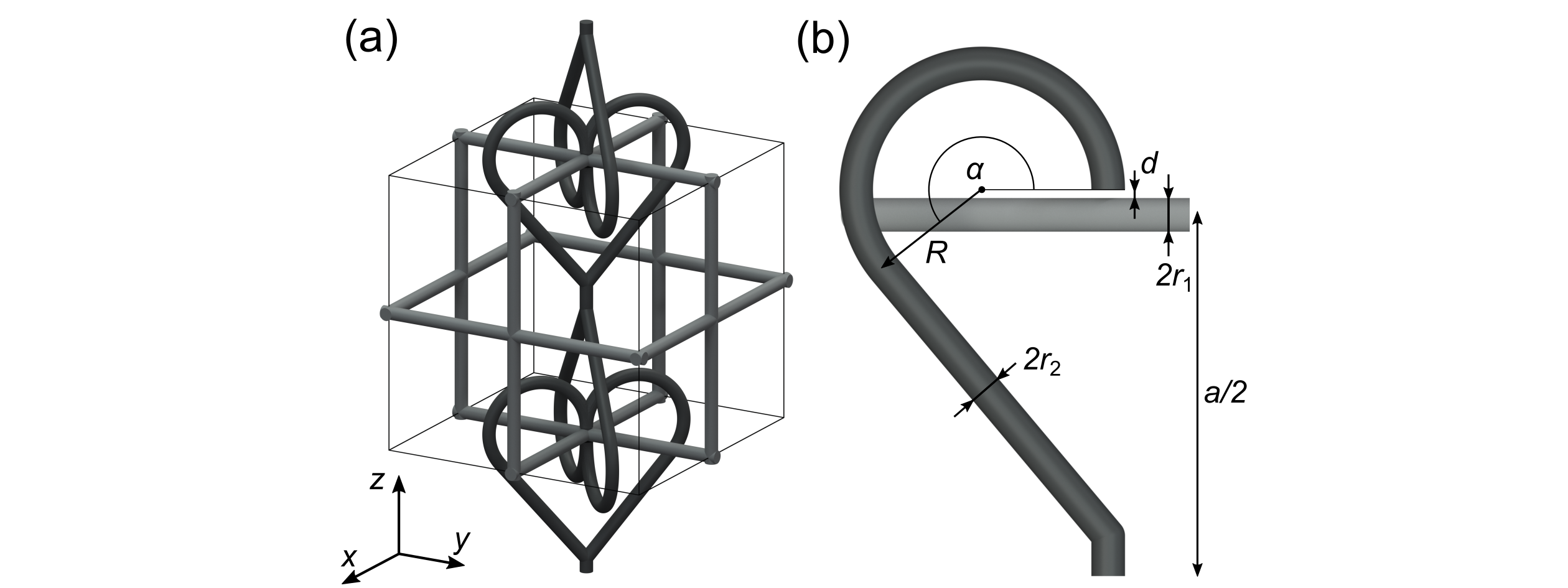}
	\caption{(a) Schematic depiction of the structure shown in figure\,16, shown in grey, with an additional structure, shown in black, yielding a local sign-inversion of the $z$-component of the magnetic field. If the magnetic permeability of the black material in vacuum is sufficiently large, {\it i.e.}, $\mu^1 \gg 1$, the structure guides the magnetic field lines such that, in the small region where the two heart-shaped elements nearly touch and where the crosses parallel to the $xy$-plane are located, the $z$-component of the magnetic field changes sign. Note the two-fold rotational symmetry about the $z$-axis. The highly-permeable elements are rotated by $45^{\circ}$ with respect to the principal axes. Combining such elements for all three axes yields the metamaterial shown in figure\,18. (b) Parametrization of the structure.}
\end{figure}

In order to flip the sign of the effective Hall coefficient, we can add an appropriate permeability distribution. This is achieved using a second structure with a large magnetic permeability $\mu^1$. Note that the crosses of the electrical structure contribute to the effect only if the component of the magnetic field perpendicular to the cross is nonzero. Intuitively, the current flowing through one of the cylinders generates, in the presence of a perpendicular magnetic field, a local Hall voltage which is picked up by the other cylinder. Hence, if we are able to invert this component of the magnetic field locally for each cross, we end up with a sign-inversion of the effective Hall coefficient. A combination of the structure shown in figure\,16 with such highly-permeable elements leading to a local inversion of the $z$-component of the magnetic field is shown in figure\,17(a). We assume that the average magnetic field is along $\hat{\bm{z}}$. The highly-permeable material guides the magnetic field lines such that in the narrow region in between the intertwined elements the magnetic field is locally reversed, \textit{i.e.}, along $-\hat{\bm{z}}$. The crosses parallel to the $xy$-plane are placed in these regions of reversed magnetic field.

\begin{figure}[h!]
	\includegraphics[width=16cm]{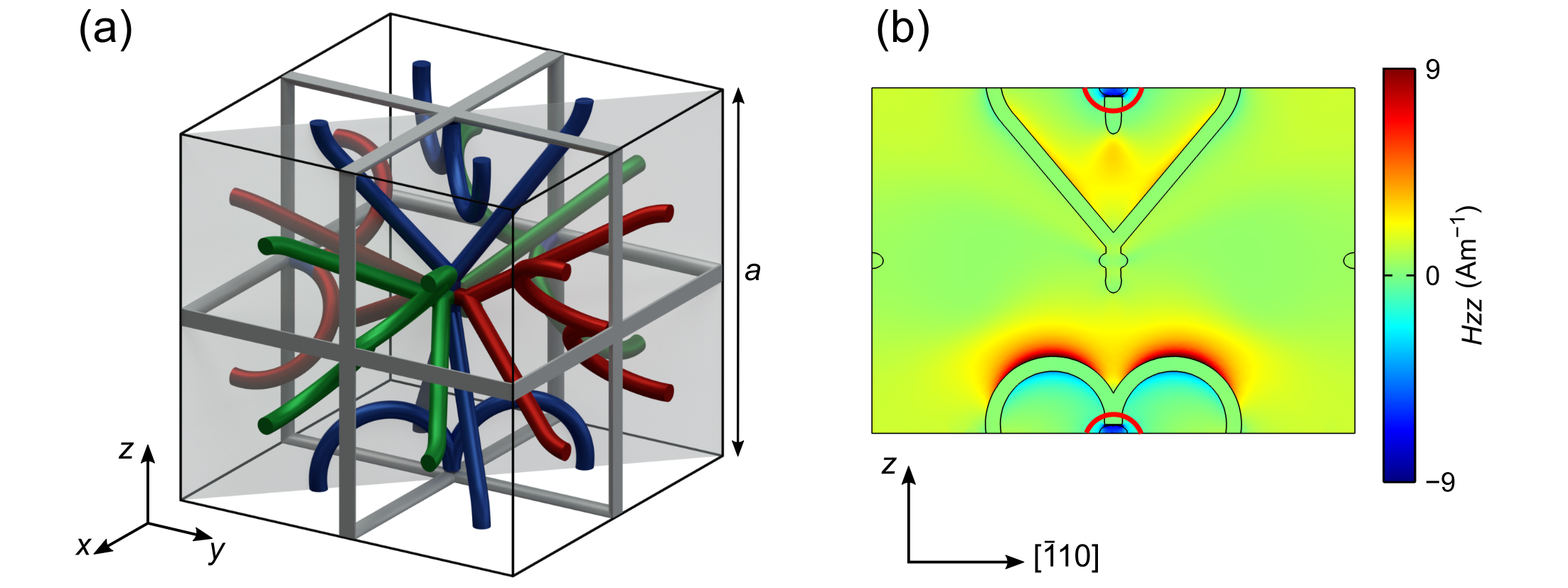}
	\caption{(a) Unit cell of a metamaterial exhibiting a sign-inversion of the Hall coefficient based on a magnetic permeability distribution. The grey parts are electrically conducting and have a finite Hall coefficient. The colored parts have a large magnetic permeability. The coloring serves for clarity only, all of these parts are made from the same constituent material. A diagonal cut through the unit cell is indicated by the grey plane. (b) Numerical calculation of $H_{zz}$, one component of the matrix-valued magnetic field, for the cut through the unit cell shown in (a). The average value of ${H}_{zz}$ is normalized. In the crucial regions of the corresponding crosses (marked by the red half-circles), $H_{zz}$ turns negative. Parameters are $R=15\,\si{\micro\meter}$, $r_1=r_2=2\,\si{\micro\meter}$, $d=0.2\,\si{\micro\meter}$, $\alpha=220^{\circ}$, $a=86\,\si{\micro\meter}$, $\sigma_0^0=200\,\si{\siemens \meter^{-1}}$, $\mu^1=1000$.}
\end{figure}

If we add such elements for all three axes, we obtain the metamaterial shown in figure\,17(a). Two things should be noted. First, the highly-permeable structure is neither connected to the semiconductor structure nor  is it connected over more than one unit cell. Therefore, it neither short-circuits nor modifies the electrical current paths in the semiconductor. Hence, we can choose the electrical conductivity and the Hall coefficient of this structure arbitrarily, \textit{e.g.}, $\sigma_{0}^{1}=A_{\text{H}}^1=0$, without affecting the effective properties. To mechanically stabilize the overall arrangement, the semiconductor rods and the magnetic rods need to be embedded in a passive material. This surrounding material has to be electrically isolating and non-magnetic ($\mu=1$). Many polymers would suit this purpose. Second, the magnetic field is only reversed for every second cross. This is sufficient to achieve a sign-inversion of the Hall coefficient, as the magnetic field is not only inverted but at the same time concentrated at the crosses. The effective electrical properties of this combined structure are isotropic by symmetry as well, since the structure has the cubic crystallographic point group $32$. A numerical calculation of $H_{zz}$, one component of the matrix-valued magnetic field $\bm{H}$, for a cut through the unit cell is shown in figure\,18(b). It can be seen that ${H}_{zz}$ turns negative at the center of the top and bottom boundaries of the unit cell where the crosses parallel to the $xy$-plane are located.

Using equations\,(\ref{constitmacr}), (\ref{magpermeff}) and (\ref{effHallmatMag}), we can calculate the effective properties. Using the parameters given in figure\,17, we obtain $\sigma_0^*=3.48\cdot 10^{-3}\sigma_0^0$, $\mu^*=3.14\cdot 10^{-3}\mu^1$, and $A_{\text{H}}^{*\mu}=-7.5 A_{\text{H}}^0$. If we assume a trivial magnetic permeability distribution (\textit{i.e.}, $\mu=1$ everywhere) instead, we obtain $A_{\text{H}}^{*0}=11.86 A_{\text{H}}^0$.

\section{Exceeding previous bounds}
As a second example, based on the structure shown in figure\,16, we employ distributions of the magnetic permeability to concentrate the magnetic field in the regions of large cofactor, leading to an increase in the effective Hall coefficient and the effective Hall mobility. 

\begin{figure}[h!]
	\includegraphics[width=7cm]{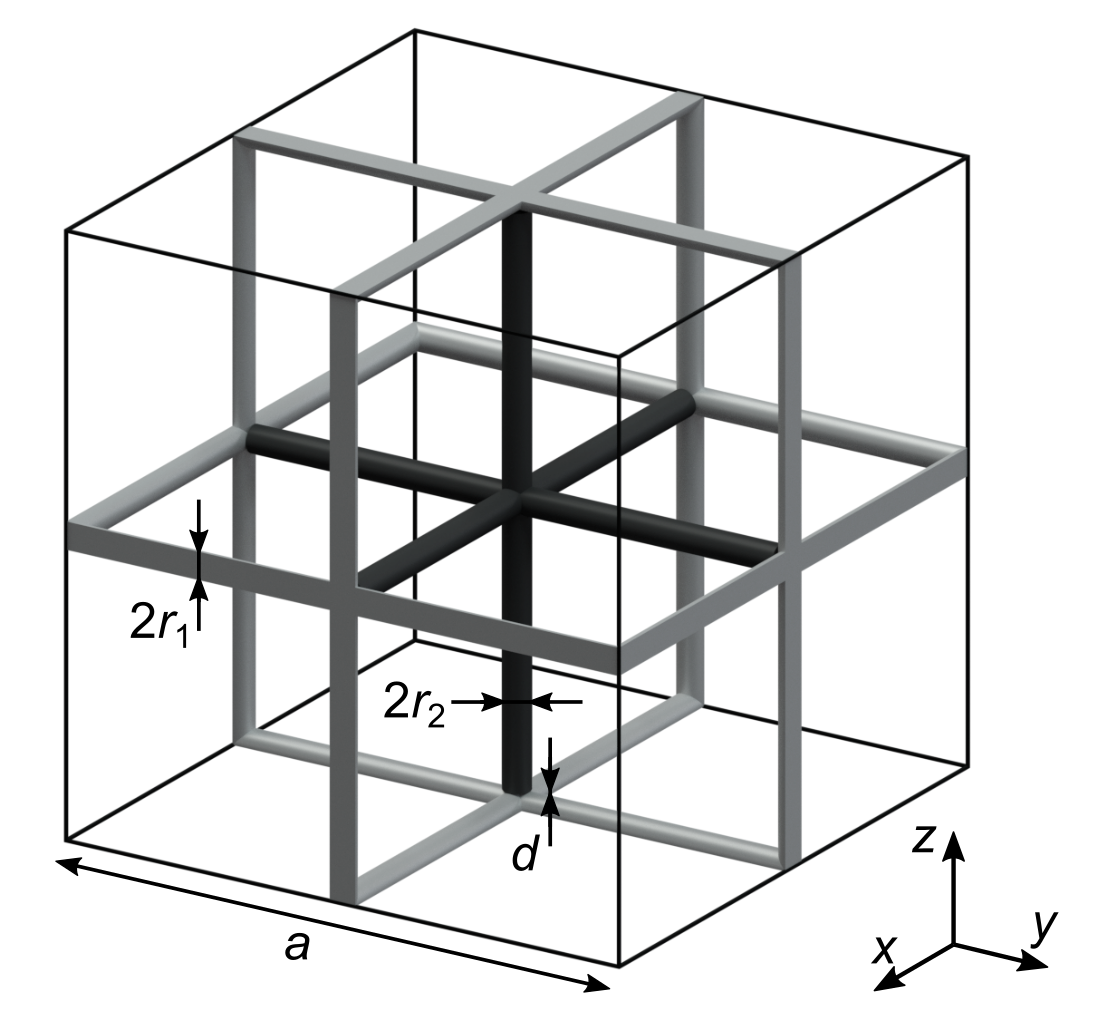}
	\caption{Metamaterial unit cell based on the electrically conducting structure shown in figure\,16, which is depicted in grey here. The black rods have a large magnetic permeability and concentrate the magnetic field at the crossing points of the electrical structure.}
\end{figure}

Clearly, the effective isotropic Hall coefficient of the arrangement of crossing electrically conductive cylinders, shown in figure\,16 and figure\,19, $A_{\text{H}}^{*0}$, is much larger than the bulk Hall coefficient of the material, $A_{\text{H}}^0$, due to current confinement. One expects the effective Hall coefficient to scale asymptotically like $a/r_1$. From the numerical calculations, we obtain $A_{\text{H}}^{*0}\approx 0.275 \left(a/r_1 \right)A_{\text{H}}^0$. Likewise, the effective conductivity, $\sigma_0^{*0}$, is much smaller than $\sigma_0^0$. Elementary considerations yield $\sigma_0^{*0}=2\pi\left(r_1/a\right)^2\sigma_0^{0}$, which is in excellent agreement with our numerical calculations (not depicted). Thus, the product $A_{\text{H}}^{*0} \sigma_0^{*0}$ is bounded and scales as $r_1/a$. 

Adding the lattice of black rods with magnetic permeability $\mu^1 \gg 1$ and radius $r_2$, as shown in figure\,19, concentrates the magnetic field at the crossing points of the electrical structure. This leads to an increased normal component of the $\bm{b}$-field at these locations, and hence, to an increased local Hall voltage. Obviously, the effective zero magnetic-field conductivity does not change, $\sigma_0^{*\mu}=\sigma_0^{*0}$. Analogously to the sign-inversion structure, only every second semiconductor crossing sees an increased magnetic flux density. Therefore, only half of the crossings contribute to the increased effective Hall coefficient. 

Again, we calculate the effective parameters by evaluating equations\,(\ref{constitmacr}), (\ref{magpermeff}) and (\ref{effHallmatMag}) using COMSOL Multiphysics. Furthermore, the effective permeability of the structure can be estimated by elementary considerations using the concept of magnetic circuits and approximately calculating the reluctance of a single unit cell. The total magnetic flux is given by the analogue of Ohm's law with a scalar magnetic potential difference given by the lattice constant, leading to a normalized average $\bm{h}$-field. In order to find an expression for the effective Hall coefficient including the magnetic permeability distribution, $A_{\text{H}}^{*\mu}$, one has to consider the field enhancement at the crossing points, which can be estimated easily.

As it turns out, there is no combination of parameters breaking the bound for the effective Hall mobility. In the limit of $d \ll r_1$, $r_1 \ll r_2$ and $r_2 \ll a$, we have that $A_{\text{H}}^{*\mu} \sigma^{*\mu}_0\rightarrow A_{\text{H}}^{0} \sigma^{0}_0$ while $A_{\text{H}}^{*\mu} \propto 1/r_1$.

\begin{figure}[h!]
	\includegraphics[width=16cm]{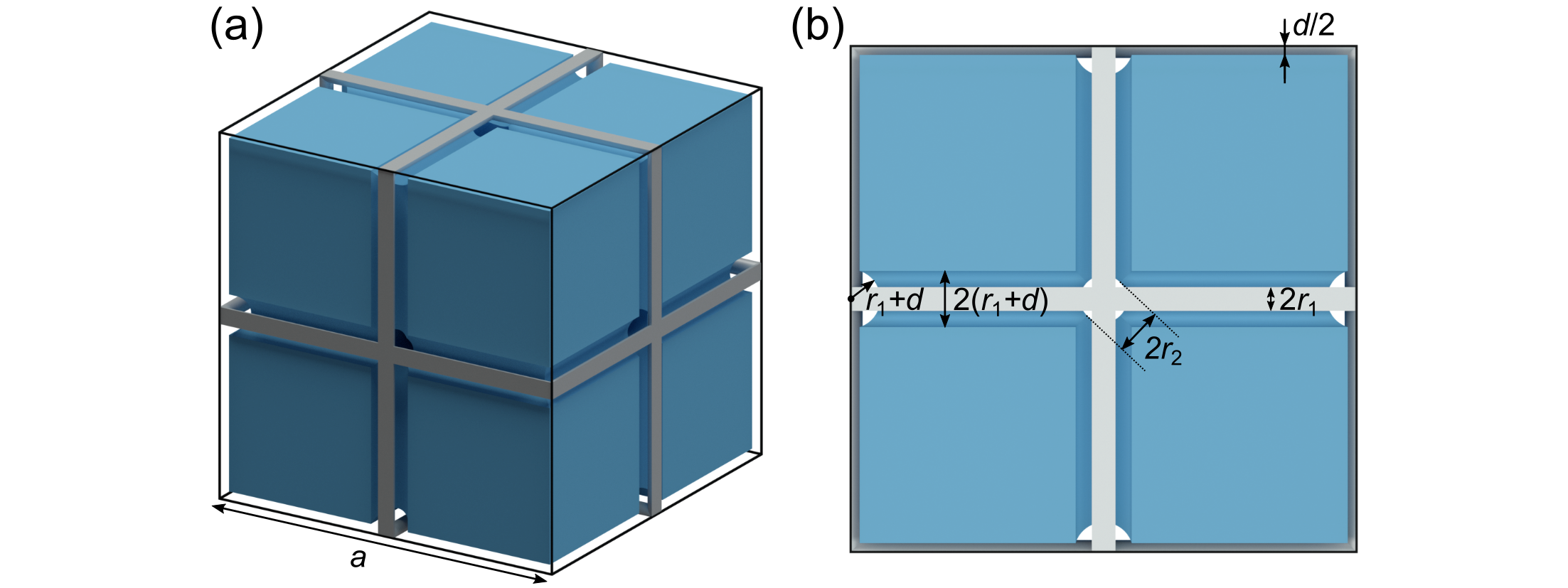}
	\caption{Unit cell of a metamaterial breaking the bound on the effective Hall mobility. The semiconductor structure shown in grey is the same as before. The material shown in blue has magnetic permeability $\mu = 0$, \textit{i.e.}, it is a superconductor. Voids in this material play the role of the highly-permeable rods and channel the magnetic field. In order to avoid electrical short-circuiting, we added gaps of $\mu=1$ in between the semiconductor and the superconductor and in between the superconducting parts of different unit cells. (b) Top view of the unit cell shown in (a).}
\end{figure}

The situation is distinct if we make the (mathematical) assumption that the regions of large permeability (the black rods) extend through the electrical structure (thereby introducing an additional constituent material and neglecting short-circuiting of the structure). In this case, the magnetic field lines are constrained to the rods. Hence, the field enhancement is given by $(a/r_2)^2/\pi$ and the bound can be broken easily.

A related, yet different possibility is illustrated in figure\,20. The grey parts correspond to the structure shown in figure 16. The magnetic field lines are guided and concentrated in the air voids with $\mu = 1$ because the field cannot penetrate into the material with zero magnetic permeability (blue). An ideal superconductor fulfills this condition of $\mu = 0$ (the Meissner effect). The combination of a small air spacing between the superconductor and the electrical parts and the fact that the superconducting parts are not connected avoids short circuits. For the parameter choice $r_1=0.5\,\si{\micro\meter}$, $r_2=2r_1$, $d=0.01\,\si{\micro\meter}$, and $a=86\,\si{\micro\meter}$, we obtain an effective Hall mobility of $A_{\text{H}}^{*\mu} \sigma^{*\mu}_0=3.85A_{\text{H}}^{0} \sigma^{0}_0$. This proves by conceptual example that the bound on the effective Hall mobility can be broken by additional distributions of the magnetic permeability---albeit merely by a factor of about two for the already rather extreme choice of parameters. Experimental realization of such structures does not appear to be in reach with current technology.

\section{Conclusion}

In conclusion, we have provided a comprehensive discussion of homogenization theory for conductive metamaterials subject to a weak static magnetic field. We consider classical transport and three-dimensional periodic (\textit{i.e.}, crystalline) passive structures.  The metamaterial unit cell is allowed to exhibit not only arbitrary spatial distributions of the conductivity and the Hall coefficient, but also of the static magnetic permeability of the constituent materials. Perhaps most importantly, we have shown that inspecting the cofactor field provides a direct and intuitive way to identify crucial parts of structures. This approach has been applied to known structures, including three-dimensional chainmail-like metamaterials composed of a cubic arrangement of interlinked tori. In addition, we have suggested and discussed two new and conceptually distinct architectures which also exhibit a sign reversal of the effective isotropic Hall coefficient. Furthermore, structures with effective highly anisotropic Hall coefficient tensors have been discussed on the same footing as well.\\[0,2cm]
\noindent
It is clear that the elements of the magnetic-field independent part of the conductivity tensor cannot be negative due to energy conservation and due the second law of thermodynamics. In addition, Onsager's principle imposes restrictions on the magnetic-field dependent part. However, it is presently not clear whether {\it any} effective conductivity and Hall coefficient tensors compatible with these fundamental constraints are actually accessible by metamaterials.  

\section*{Acknowledgments}
We acknowledge support by the Karlsruhe School of Optics \& Photonics (KSOP), the Hector Fellow Academy, the KIT Nanostructure Service Laboratory (NSL), and the Helmholtz Program Science and Technology of Nanosystems (STN). G\,W  Milton is grateful for support from the National Science Foundation through Grant NSF-1211359. G\,W  Milton and M Kadic  acknowledge support by the Labex ACTION program (Contract No. ANR-11-LABX-0001-01). M Kadic was supported by the French ``Investissements d'Avenir'' program, project ISITE-BFC (Contract No. ANR-15-IDEX-03). We thank Marc Briane (INSA de Rennes) and Alexander M\"unchinger (KIT) for helpful discussions. We acknowledge support by Deutsche Forschungsgemeinschaft and Open Access Publishing Fund of Karlsruhe Institute of Technology.

\section*{References}
\providecommand{\newblock}{}


\begin{thebibliography}{10}
	\expandafter\ifx\csname url\endcsname\relax
	\def\url#1{{\tt #1}}\fi
	\expandafter\ifx\csname urlprefix\endcsname\relax\def\urlprefix{URL }\fi
	\providecommand{\eprint}[2][]{\url{#2}}
	% Bibliography created with iopart-num v2.1
	% /biblio/bibtex/contrib/iopart-num
	
	\bibitem{Yu05}
	Yu P and Cardona M 2010 {\em Fundamentals of Semiconductors\/} (Berlin,
	Heidelberg: Springer-Verlag)
	
	\bibitem{Popovic04}
	Popovic R\ S 2004 {\em Hall Effect Devices\/} (Bristol, Philadelphia: Institute
	of Physics Publishing)
	
	\bibitem{QHE}
	von Klitzing K, Dorda G and Pepper M 1980 {\em Phys. Rev. Lett.\/} {\bf 45}
	494--497
	
	\bibitem{Soukoulis11}
	Soukoulis C~M and Wegener M 2011 {\em Nat. Photonics\/} {\bf 5} 523--530
	
	\bibitem{Bauer17}
	Bauer J, Meza L~R, Schaedler T~A, Schwaiger R, Zheng X and Valdevit L 2017 {\em
		Adv. Mater.\/} {\bf 29}
	
	\bibitem{Cummer16}
	Cummer S~A, Christensen J and Al\`u A 2016 {\em Nat. Rev. Mater.\/} {\bf 1}
	16001
	
	\bibitem{Alu14}
	Al{\`u} A 2014 {\em Physics\/} {\bf 7} 12
	
	\bibitem{Pendry99}
	Pendry J\ B, Holden A\ J, Robbins D\ J and Stewart W\ J 1999 {\em IEEE Trans.
		Microw. Theory Tech.\/} {\bf 47} 2075--2084
	
	\bibitem{Soukoulis07}
	Soukoulis C~M, Linden S and Wegener M 2007 {\em Science\/} {\bf 315} 47--49
	
	\bibitem{Mei06}
	Mei J, Liu Z, Wen W and Sheng P 2006 {\em Phys. Rev. Lett.\/} {\bf 96} 024301
	
	\bibitem{Fang06}
	Fang N, Xi D, Xu J, Ambati M, Srituravanich W, Sun C and Zhang X 2006 {\em Nat.
		Mater.\/} {\bf 5} 452--456
	
	\bibitem{Gatt08}
	Gatt R and Grima J~N 2008 {\em Phys. Stat. Sol. RRL\/} {\bf 2} 236--238
	
	\bibitem{Qu17}
	Qu J, Kadic M and Wegener M 2017 {\em Appl. Phys. Lett.\/} {\bf 110} 171901
	
	\bibitem{Ros05}
	Ros A, Eichhorn R, Regtmeier J, Duong T~T, Reimann P and Anselmetti D 2005 {\em
		Nature\/} {\bf 436} 928--928
	
	\bibitem{Yang13}
	Yang F, Mei Z~L, Yang X~Y, Jin T~Y and Cui T~J 2013 {\em Adv. Funct. Mater.\/}
	{\bf 23} 4306--4310
	
	\bibitem{Milton02}
	Milton G~W 2002 {\em The Theory of Composites\/} (Cambridge: Cambridge
	University Press)
	
	\bibitem{Briane09}
	Briane M and Milton G~W 2009 {\em Arch. Rational Mech. Anal.\/} {\bf 193}
	715--736
	
	\bibitem{Briane08}
	Briane M, Manceau D and G\ W~M 2008 {\em J. Math. Anal. Appl.\/} {\bf 339} 1468
	-- 1484
	
	\bibitem{Kern17}
	Kern C, Kadic M and Wegener M 2017 {\em Phys. Rev. Lett.\/} {\bf 118} 016601
	
	\bibitem{Briane10}
	Briane M and Milton G~W 2010 {\em SIAM J. Appl. Math.\/} {\bf 70} 1810--1820
	
	\bibitem{Kern15}
	Kern C, Kadic M and Wegener M 2015 {\em Appl. Phys. Lett.\/} {\bf 107} 132103
	
	\bibitem{Kern17-2}
	Kern C, Schuster V, Kadic M and Wegener M 2017 {\em Phys. Rev. Applied\/} {\bf
		7} 044001
	
	\bibitem{Bensoussan78}
	Bensoussan A, Lions J~L and Papanicolaou G 1978 {\em Asymptotic analysis for
		periodic structures\/} (Amsterdam: North-Holland Publishing Company)
	
	\bibitem{Bergman83}
	Bergman D\ J 1983 Self-duality and the low field hall effect in 2d and 3d
	metal-insulator composites {\em Percolation Structures and Processes\/} ed
	Deutscher G, Zallen R and Adler J (Jerusalem: Israel Physical Society) pp
	297--321
	
	\bibitem{Mani17}
	Mani R\ G 2017 {\em Phys. Today\/} {\bf 70}(7) 13
	
	\bibitem{Oswald18}
	Oswald J 2018 {\em Phys. Rev. Lett.\/} {\bf 120} 149701
	
	\bibitem{Wegener17}
	Wegener M, Kadic M and Kern C 2017 {\em Phys. Today\/} {\bf 70}(10) 14
	
	\bibitem{Kern18}
	Kern C, Kadic M and Wegener M 2018 {\em Phys. Rev. Lett.\/} {\bf 120} 149702
	
	\bibitem{Onsager31}
	Onsager L 1931 {\em Phys. Rev.\/} {\bf 37} 405--426
	
	\bibitem{Onsager31_2}
	Onsager L 1931 {\em Phys. Rev.\/} {\bf 38} 2265--2279
	
	\bibitem{LandauStat}
	Landau L\ D and Lifshitz E\ M 1980 {\em Statistical Physics, Part 1\/} (Oxford:
	Pergamon Press) \S 120
	
	\bibitem{LandauCont}
	Landau L\ D and Lifshitz E\ M 1984 {\em Electrodynamics of Continuous Media\/}
	(Oxford: Pergamon Press) \S22
	
	\bibitem{Murat97}
	Murat F and Tartar L 1997 H-convergence {\em Topics in the Mathematical
		Modelling of Composite Materials\/} ed Cherkaev A and Kohn R (Boston:
	Birkh{\"a}user Boston) pp 21--43
	
	\bibitem{Bergman94}
	Bergman D~J and Strelniker Y~M 1994 {\em Phys. Rev. B\/} {\bf 49} 16256--16268
	
	\bibitem{Tornow96}
	Tornow M, Weiss D, Klitzing K~v, Eberl K, Bergman D~J and Strelniker Y~M 1996
	{\em Phys. Rev. Lett.\/} {\bf 77} 147--150
	
	\bibitem{Bergman17}
	Strelniker Y~M and Bergman D~J 2017 {\em Phys. Rev. B\/} {\bf 96} 235308
	
	\bibitem{camareddine01}
	Camar-Eddine M and Seppecher P 2001 {\em C. R. Acad. Sci. - Series I -
		Mathematics\/} {\bf 332} 485 -- 490
	
	\bibitem{camareddine02}
	Camar-Eddine M and Seppecher P 2002 {\em Math. Models Meth. Appl. Sci.\/} {\bf
		12} 1153--1176
	
	\bibitem{Kadic15}
	Kadic M, Schittny R, B\"uckmann T, Kern C and Wegener M 2015 {\em Phys. Rev.
		X\/} {\bf 5} 021030
	
	\bibitem{Briane04}
	Briane M, Milton G\ W and Nesi V 2004 {\em Arch. Rational Mech. Anal.\/} {\bf
		173} 133--150
	
	\bibitem{Alessandrini01}
	Alessandrini G and Nesi V 2001 {\em Arch. Rational Mech. Anal.\/} {\bf 158}
	155--171
	
	\bibitem{Mani94}
	Mani R~G and von Klitzing K 1994 {\em Appl. Phys. Lett.\/} {\bf 64} 1262--1264
	
	\bibitem{Briane10-2}
	Briane M and Milton G~W 2009 {\em Multiscale Model. Simul.\/} {\bf 7}
	1405--1427
	
\end{thebibliography}
\end{document}